\documentclass[12pt,a4paper]{article}
\input epsf
\usepackage{amsmath,amsfonts,epsfig,color,latexsym}
\usepackage{amssymb}
\usepackage[english, USenglish]{babel}
\usepackage{epsfig}
\usepackage{a4wide}
\usepackage{rotating}

\usepackage[numbers,square,comma, compress]{natbib} 
\usepackage{bm}
\usepackage{tabularx}
\usepackage{booktabs}

\newcommand{\N}{\mathcal{N}}

\csname @addtoreset\endcsname{equation}{section}

\newcommand{\be}{\begin{equation}}
\newcommand{\ee}{\end{equation}}
\newcommand{\bea}{\begin{eqnarray}}
\newcommand{\eaa}{\end{eqnarray}}

\newcommand{\pa}{\partial}

\newcommand{\cN}{{\cal N}}

\newcommand{\bt}[1]{{\bar t}}

\renewcommand{\theequation}{\thesection.\arabic{equation}}

\newcommand{\fun}{\mathbb{F}}

\newcommand{\amp}{\mathcal{A}}

\csname @addtoreset\endcsname{equation}{section}
\makeatletter

\renewcommand{\@fnsymbol}[1]{\@alph{#1}}
\makeatother
\newcommand{\comment}[1]{}
\def\Gc{{\cal G}}
\def\Hc{{\cal H}}
\def\Fc{{\cal F}}
\def\Nc{{\cal N}}\def\Mc{{\cal M}}
\def\fc#1#2{{\frac{#1}{#2}}}
\newcommand{\req}[1]{(\ref{#1})}
\def\T{{\mathbb{T}}}
\def\Si{\Sigma}
\def\Oc{{\cal O}}
\def\Jc{{\cal J}}

\def\ov{\overline}

\author{\\[-0.3cm]\large S.~Hohenegger\thanks{{\tt shoheneg@mppmu.mpg.de}}\ \  \ 
and\ \ S.~Stieberger\thanks{{\tt stephan.stieberger@mpp.mpg.de}}}
\title{\begin{flushright}{\vspace{-0.8cm}\normalsize MPP--2011--77}\end{flushright}
\vspace{0.8cm}
\bf{BPS Saturated String Amplitudes:\\[2mm] 
$\bm{K3}$ Elliptic Genus  and Igusa Cusp Form $\bm{\chi_{10}}$}\\[0.5cm]}
\dateUSenglish
\date{}
\graphicspath{{figures/}}

\begin{document}

\begin{titlepage}

\maketitle
\begin{center}
\renewcommand{\thefootnote}{\fnsymbol{footnote}}\vspace{-1cm}
\it Max--Planck--Institut f\"ur Physik\\ 
\it  Werner--Heisenberg--Institut\\
\it  80805 M\"unchen, Germany\\[2.5cm]
\end{center}
\begin{abstract}\baselineskip20pt
We study BPS saturated one--loop amplitudes in type II string theory
compactified on $K3\times \T^2$. The classes of amplitudes we consider
are only sensitive to the very basic topological data of the internal
$K3$ manifold. As a consequence, the integrands of the former are
related to the elliptic genus of $K3$, which can be decomposed into
representations of the internal $\cN=4$ superconformal
algebra. Depending on the precise choice of external states these
amplitudes capture either only the contribution of the short multiplets or the full series including intermediate multiplets. In the latter case we can define a generating functional for the whole class, which we show is given by the weight ten Igusa cusp form 
$\chi_{10}$ of $Sp(4,\mathbb{Z})$. We speculate on possible algebraic implications of our result on the BPS states of the $\N=4$ type~II compactification. 
\end{abstract}
\thispagestyle{empty}
\end{titlepage}

\tableofcontents

\vspace{.0in}



\vskip3cm

\section{Introduction}\baselineskip16pt
BPS--saturated amplitudes compute a very peculiar type of couplings in the effective action of string theory with extended supersymmetry. They receive contributions only from a particular class of states in the full Hilbert space, which are annihilated by a subset of the supercharges. As a consequence, these amplitudes have very interesting and unique properties. In the recent years, many examples of such couplings in string theories with $\N=2$ supersymmetry \cite{Antoniadis:1993ze,Antoniadis:1995zn,Antoniadis:1996qg,Morales:1996bp,Antoniadis:2009nv,AHNS} as well as $\N=4$ supersymmetry \cite{Berkovits:1994vy,Lerche:1999ju,Hammou:1999in,Antoniadis:2006mr,Antoniadis:2007cw,Antoniadis:2009tr} have been discovered. These amplitudes yield valuable information about the internal manifold of the underlying string compactification, which is encoded through a very particular dependence on the moduli. A closer study reveals, that the latter can be captured in the form of differential equations \cite{Bershadsky:1993cx,Antoniadis:2007cw,Antoniadis:2009nv,AHNS}, which in many cases \cite{Grimm:2007tm,Huang:2006si}  allow to even compute explicit expressions for the string amplitudes. 

Moreover, BPS saturated objects in string theory also teach us interesting lessons on algebraic aspects of the space of BPS states. Indeed, in \cite{Harvey:1995fq,Harvey:1996gc} (see also \cite{Moore:1997ar} for an overview) Harvey and Moore argue, that the space of BPS--states in string theory forms an algebra. By studying certain 
one--loop corrections in heterotic $\N=2$ compactifications and relating them to the denominator formula of a (generalized) Borcherds--Kac-Moody (BKM) algebra \cite{Borcherds0} they have obtained further valuable hints on the nature of this 'algebra of BPS states'. Similar results for the $E_8\times E_8$ heterotic string compactified on $\T^2$ have recently been obtained in \cite{Gaberdiel:2011qu}. There it is shown, that the space of BPS--states forms a representation of a BKM--algebra (which is constructed explicitly). The denominator formula of an extension of the latter appears in a certain heterotic one--loop $\mathcal{N}=4$ topological string amplitude, which has previously been studied in \cite{Antoniadis:2006mr,Antoniadis:2007cw}. A generalization of these results to different gauge groups is studied in \cite{HP}. Since this approach is fully perturbative the topological amplitude studied in this way is only sensitive to perturbative BPS states (\emph{i.e.} $1/2$ BPS states after compactification down to four-dimensions).

In this work we  investigate the possibility of BPS saturated amplitudes, which are also sensitive to $1/4$ BPS states in a theory with $\N=4$ supersymmetry. To this end, we shall consider one--loop amplitudes in type II string theory compactified on $K3\times \mathbb{T}^2$ and we shall particularly be interested in quantities, which have a certain 'index-like' behaviour w.r.t. the internal CFT living on $K3$, \emph{i.e.} we want the amplitude to be invariant under particular deformations of the latter (see especially \cite{Cecotti:1992qh} for an analogue in $\N=2$). Indeed, such an object would have to be sensitive only to very basic topological data of $K3$ and a natural candidate in the $\cN=4$ setup is the elliptic genus of $K3$ \cite{SW,Witten:1986bf,Lerche:1987qk}. The latter has recently attracted a lot of interest following the observation, that it might carry an action of the Mathieu group $\mathbb{M}_{24}$ \cite{Eguchi:2010ej,Cheng:2010pq,Gaberdiel:2010ch,Eguchi:2010fg}. Its appearance in the form of a BPS saturated amplitude might therefore give additional hints about algebraic properties of the space of $\N=4$ BPS states. Besides this, the elliptic genus is also the seed function for an infinite product representation of a certain weight ten Siegel modular form of $Sp(4,\mathbb{Z})$, known as the Igusa cusp form $\chi_{10}$. The latter is proposed  to encode the degeneracies of (non-perturbative) dyonic $1/4$ BPS-states in $\mathcal{N}=4$ string compactifications \cite{Dijkgraaf:1996it}. It has already been shown earlier in \cite{GritsenkoNikulin}, that this infinite product representation can be related to the denominator formula of a certain rank 3 (super) BKM-algebra. This suggests that degeneracies of dyons also become related to the root multiplicities of the associated BKM--algebra. The physical role of this BKM--algebra has been further clarified in \cite{Cheng:2008fc} (see also \cite{Cheng:2008gx,Govindarajan:2008vi,Govindarajan:2009qt}), where it has been shown, that the wall--crossing behaviour of the dyon spectrum is controlled
by the hyperbolic Weyl group of this BKM--algebra.

A series of $1/4$ BPS saturated one--loop amplitudes has  been studied in \cite{Lerche:1999ju}. In this reference the (fermionic contractions of the)  $2K+4$--point couplings $(\partial_\nu T\partial^\nu U)^{K+1}(\partial_\rho\phi\partial^{\rho}\bar{\phi})$ are computed for $K\geq 0$ at the one-loop level in type II superstring theory compactified on $K3\times \T^2$. Here $\phi\in\{T,U\}$ describe the K\"ahler and complex structure moduli of the  torus $\T^2$. However, this set of couplings does not entail the elliptic genus of $K3$, but rather another index--like object related to the torus $\T^2$. Indeed, the only topological information on $K3$ entering these particular couplings is the Euler number. It is one of the motivations for this work to investigate  amplitudes, which encode slightly more topological information on $K3$. Indeed, we propose several new BPS--saturated amplitudes, which we show to be related to (parts of) the elliptic genus of $K3$. In fact, we can distinguish between two different scenarios: In a first approach we consider a coupling of the form $R_{(+)}^2 F_{(-)}^{2N}$, where $R_{(+)}$ is the self-dual part of the Riemann tensor and $F_{(-)}$ the anti-self-dual part of the field strength tensor of one of the Kaluza-Klein vector fields stemming from the compactification on $K3\times \T^2$. In particular, all these component fields are part of (massless) $1/2$ BPS short multiplets in the $\cN=4$ string effective action. We show, that this coupling can be linked to the contribution of massless short $\cN=4$ multiplets to the elliptic genus of $K3$. In fact, after summing over $N$, we recover a particular Appell--Lerch sum, which falls into the class of \emph{mock modular forms}, that have been introduced by Ramanujan and which have recently attracted a lot of attention both in mathematics and physics (see e.g. \cite{Zagier2} for a nice overview). We also discuss the exact structure of these effective action couplings using $\cN=4$ harmonic superspace in the supergravity frame. In order to obtain the full elliptic genus (including also the contribution of the intermediate multiplets), we have to consider amplitudes, whose external fields are part of massive multiplets. We compute those amplitudes in two distinct ways: First as a reducible limit of a $1/2$ BPS amplitude (\emph{i.e.} with the position of two of the external insertions colliding) and secondly directly using massive vertices. Upon introducing a coupling constant $\lambda $, we demonstrate in both ways that we can indeed define a generating functional of one--loop string amplitudes, which eventually resembles the elliptic genus of $K3$.  We explicitly perform also the world-sheet torus integral and prove that this generating functional is given by the weight ten Igusa cuspform $\chi_{10}$ of $Sp(4,\mathbb{Z})$.

This work is organized as follows: In section~\ref{SPEL} we will
review important aspects of the BPS spectrum of type IIA string
theory compactified on $K3\times \T^2$, emphasizing important
BPS-saturated index-like objects. In section \ref{Sect:12BPSandMock}
we will first consider a one-loop amplitude with external fields
residing in short ($1/2$ BPS) multiplets. We will evaluate this amplitude up to the integration over the world-sheet torus and find that the integrand is related to a particular part of the elliptic genus of $K3$. Indeed, by expanding the latter in terms of characters of $\N=4$ representations (along the lines of \cite{Eguchi:1988vra,Eguchi:2009cq,Eguchi:2010ej}) we see that the integrand is given by a particular Appell-Lerch sum, which captures the contributions from short $\N=4$ multiplets. We give also a superspace description of this particular class of couplings in terms of harmonic superspace in section~\ref{Sect:SuperspaceGrav}. In a next step in section~\ref{Sect:14BPSamps} we consider a class of amplitudes which contain massive external fields. We provide two ways of computing these amplitudes: $(i)$ as collinear limits of $1/2$ BPS amplitudes; $(ii)$ directly by the use of massive scalar field vertices. Both methods lead to the same result, a torus integral, whose integrand is given by derivatives of the (full) elliptic genus of $K3$. In section~\ref{Sect:LoopIntegral}, by introducing a coupling constant $\lambda $, we define a generating functional for this class of amplitudes. Direct evaluation of the torus integration shows that this functional is the Igusa cusp form $\chi_{10}$, which depends on $\lambda $ as well as the moduli of the internal $\T^2$ torus. This work is accompanied by five appendices, which contain additional information about the internal CFT of type II $\N=4$ compactifications as well as several computations that we deemed too lengthy to fit into the main body of this paper.
\section{BPS Spectrum of Type II String Theory on $\bm{K3\times \T^2}$}\label{SPEL}

The most basic string theories with $\N=4$ supersymmetry in $D=4$ space--time dimensions
are type IIA or type IIB compactified on $K3\times \T^2$ or equivalently heterotic string theory compactified on a six--torus $\T^6$.
In the following we shall concentrate on type IIA on $K3\times \T^2$ and present its 
spectrum.

\subsection[Spectrum and BPS States of Type II on $K3\times \T^2$]{Spectrum and BPS States of Type II on $\bm{K3\times \T^2}$}\label{SPEC}

The effective  action of type II superstring theory compactified on $K3\times \T^2$  
is described by $\Nc=4$ supergravity in $D=4$. The massless spectrum consists of the $\Nc=4$ supergravity multiplet coupled to $28$ vector multiplets (VMs). As reviewed in appendix~\ref{App:LinOnShellMultiplets}, the latter contain a vector field $A_\mu$ (transforming in the adjoint of the gauge group), four Weyl spinors $\lambda^i$ and six real scalars $\varphi^{ij}$, with $\varphi^{ij}=-\varphi^{ji}$, where $i,j= 1, \ldots, 4$ are indices of the $SU(4)$ R-symmetry group.  Out of the $28$ VMs only $22$ are physical, while the remaining $6$ act as compensating multiplets. As was explained in detail in \emph{e.g.} \cite{Antoniadis:2007cw}, the $36$ scalars of these multiplets are eliminated by imposing the D-term constraints (20 constraints) as well as gauge fixing Weyl invariance (one constraint) and local SO(6) symmetry (15 constraints). The remaining $134$ scalars\footnote{Physically, in type IIA these scalars arise as follows: The 
$\sigma$--model of $K3$ has $80$ and that of $\T^2$ four real deformations. The  R--R  $1$--form gives rise to the two real scalars $C_4,C_5$ and the R--R  $3$--form gives $b_3(K3\times\T^2)=44$ scalars. Reducing the R--R $3$--form down to an anti--symmetric space--time $2$--tensor, which can be dualized to a scalar,
gives $b_1(K3\times\T^2)=2$  more scalars. Together with the dilaton field $S$ we obtain
$80+4+2+44+2+2=134$ real scalars.} from the Weyl multiplet and the remaining $22$ VMs span the coset space $\Mc$:
\be
\Mc=\fc{SU(1,1)}{U(1)}\otimes \fc{SO(22,6,\mathbb{R})}{SO(22,\mathbb{R})\times SO(6,\mathbb{R})}\ .
\label{MODSPACE}
\ee
The first factor of \req{MODSPACE} is described by the K\"ahler modulus $T=T_1+iT_2$ 
of the two--torus $\T^2$,
while its complex structure modulus $U=U_1+iU_2$, the $\sigma$--model moduli of $K3$, the type IIA dilaton field $S$ and the Wilson lines on $\T^2$ of the Ramond--Ramond (R--R) 
gauge fields parameterize the second factor. 
In type II compactification on $K3\times \T^2$ two supercharges from the left-- and right--movers each comprise the full $\N=4$ SUSY algebra. Therefore, half of the gauginos $\lambda^a$ originate from the R--NS sector and the second half $\tilde\lambda^a$ from the NS--R sector. The corresponding string world--sheet emission vertex operators of these fields are given in appendix \ref{App:Vertex}.

In type IIA the $22$ gauge vectors  $A^a_\mu$ in the VMs arise  from the R--R $3$--form potentials $C_{\mu i j}$ reduced on the $b_2(K3)=22$ two--cycles of $K3$ with the index $a$ labeling the internal SCFT, cf. appendix \ref{App:Vertex}.
On the other hand, the six graviphoton fields $B^b_\mu$  from the supergravity multiplet
stem from the R--R $1$--form $C_\mu$ in $D=10$, the R--R $3$--form $C_{\mu 45}$ reduced 
on $\T^2$, the  NS--NS anti--symmetric tensor $B_{\mu4},B_{\mu5}$ and metric 
$g_{\mu4},g_{\mu5}$ reduced along  the two one--cycles of the torus $\T^2$. 

To all these gauge fields electric and magnetic charged objects are associated.
In particular, the fundamental string wrapped on $\T^2$  with winding numbers 
$n_1,n_2$ and
Kaluza--Klein (KK) momenta $m_1,m_2$ is electrically charged under 
the gauge fields $B_{\mu4},B_{\mu5}, g_{\mu4},g_{\mu5}$ with the charges 
$n_1,n_2,m_1,m_2\in\mathbb{Z}$, respectively.
Their corresponding magnetic counter parts are  described by the NS five--brane 
wrapped on $K_3\times \mathbb{S}^1$ and a KK monopole on $\mathbb{S}^1$.
The remaining charges are carried by $D0, D2, D4$ and $D6$--branes wrapped around 
the corresponding cycles. We refer the reader to Ref. \cite{Dabholkar:2005dt} 
for a recent exhibition on these states.

The mass of a fundamental type II NS--NS closed string state wrapped around the torus 
$\T^2$ with windings $n_1,n_2$ and momenta $m_1,m_2$ is given by 
\bea
m_L^2&=&N_L-\tfrac{1}{2}+|P_L|^2+(p_\mu)^2\ ,\nonumber \\
m_R^2&=&N_R-\tfrac{1}{2}+|P_R|^2+(p_\mu)^2\ ,\label{MASS}
\eaa
with the Narain momenta $(P_L,P_R)\in \Gamma^{2,2}$
\bea
P_L&=&\frac{1}{\sqrt{2T_2U_2}}\,(m_1+m_2U+n_1\bar{T}+n_2\bar{T}U)\,,\nonumber \\
P_R&=&\frac{1}{\sqrt{2T_2U_2}}\,(m_1+m_2U+n_1T+n_2TU)\,,\label{NarainMomenta}
\eaa
corresponding to the torus $\T^2$. Level matching requires:
\begin{align}
m_L^2-m_R^2=N_L-N_R+|P_L|^2-|P_R|^2=N_L-N_R+2\ (m_1n_2-n_1m_2)=0\ .\label{LEVEL}
\end{align}
A particular interesting class of states, which will also attract our attention in the sequel,
are the so--called Dabholkar--Harvey (DH) states. The latter are perturbative BPS states, 
which may either have left--moving $N_L$ or right--moving $N_R$ excitations:
\be
\begin{array}{rrcr}
N_R,N_L&=0:\ \ \ \ \  & m_1n_2-n_1m_2=0\ , &\ \ \ \ \   1/2\ \mbox{BPS}\ ,\\[2mm]
N_L&=0:\ \ \ \ \   & 2\ (m_1n_2-n_1m_2)=N_R\ , &\ \ \ \ \   1/4\ \mbox{BPS}\ ,\\[2mm]
N_R&=0:\ \ \ \ \   & 2\ (n_1m_2-m_1n_2)=N_L\ , &\ \ \ \ \   1/4\ \mbox{BPS}\ .
\end{array}
\ee
Depending on the value of the duality invariant $m_1n_2-n_1m_2$
these states represent either $1/2$ or $1/4$ BPS saturated string states. In the remainder of this paper we will be very interested in string theory amplitudes which only receive contributions from such states. However, before explicitly computing such amplitudes, let us first discuss such BPS saturated objects from a (world-sheet) CFT point of view.

\subsection{BPS Saturated Objects in CFT}\label{App:EllGenK3}

In two-dimensional theories with extended superconformal symmetry (\emph{i.e.} with $\N\geq 2$) the study of objects, which receive contributions only from (a subset of) the BPS states has proven  to be very fruitful both from a mathematical and physical point of view. In the sequel we generically refer to such objects as being \emph{BPS saturated}. The simplest examples are (particular cases of) \emph{helicity supertraces} \cite{Kiritsis:1997gu} which are generically defined as
\begin{align}
&B_n:=\frac{1}{(2\pi i)^n}\left(\frac{\partial}{\partial z}+\frac{\partial}{\partial \bar{z}}\right)^n Z(z,\bar{z})\ \bigg|_{z=\bar{z}=0}\,,&&\text{for} &&n\in\mathbb{N}_{\text{even}}\,,\label{HelStrDef}
\end{align}
where $Z(z,\bar{z})$ is the following generating functional
\begin{align}
Z(z,\bar{z}):=\text{Tr}_{RR}\left[(-1)^{F+\bar{F}} e^{2\pi i z J_0} e^{-2\pi i \bar{z}\bar{J}_0} q^{L_0-\frac{c}{24}}\bar{q}^{\bar{L}_0-\frac{\bar{c}}{24}}\right]\,,&&\text{with} &&q:=e^{2\pi i\tau}\,.\label{GenFuncHelTr}
\end{align}
Here, $c$ is the central charge of the CFT, $F+\bar{F}$ is the total
fermion number, $L_0$ and $\bar{L}_0$ are the zero modes of the
Virasoro generators, $J_0$ and $\bar{J}_0$ the zero modes of the
(left- and right-moving) $U(1)$ R--symmetry currents and the trace is
over the full Ramond sector of the theory. We note that $B_n=0$ for
$n$ odd and for $n<\N$. In fact, for $\N\leq n< 2\N$ the only
contribution to $B_n$  stems from BPS representations (\emph{i.e.}
'short' or 'intermediate' multiplets) in the trace in
(\ref{GenFuncHelTr}), while for $n\geq2\N$ generic representations
contribute \cite{Kiritsis:1997gu}. In the case of
$\N=4$ supersymmetry, relevant in this work, the interesting
(BPS--saturated) traces are $B_4$ and $B_6$. The object $B_4$ is only sensitive to $1/2$ BPS representations
('short' multiplets) while $1/4$ BPS representations ('intermediate
multiplets') could be relevant\footnote{Indices counting M$2$--branes in $K3\times\T^2$ have been discussed in \cite{KKV} from an M--theory perspective.} for $B_6$. However, a more careful
analysis shows, that also $B_6$ only receives contributions from $1/2$
BPS representations and that the intermediate multiplets mutually cancel out. This can \emph{e.g.} be immediately seen in the case of heterotic string theory compactified on $\T^6$ (which is dual to type II on $K3\times \T^2$)  \cite{Dabholkar:2005dt,Dabholkar:2009kd}. 

A different BPS saturated quantity, which is defined for any SCFT with $\N\geq 2$ is the \emph{elliptic genus}\footnote{The elliptic genus plays an important role in the  computation of anomaly cancellation terms   
and other exact quantities in the string effective action at one--loop 
\cite{Lerche:1988zy,Lerche:1988np}.} \cite{SW,Witten:1986bf,Lerche:1987qk}
\begin{align}
\phi(\tau,z):=\text{Tr}_{RR}\left[(-1)^{F+\bar{F}}e^{2\pi izJ_0}q^{L_0-\frac{c}{24}}\bar{q}^{\bar{L}_0-\frac{\bar{c}}{24}}\right]\,,\label{EllGenDef}
\end{align}
where the trace is taken over the Ramond-sector of the CFT. Due to the usual index type of arguments only the ground states  contribute from the right moving sector and thus $\phi(\tau,z)$ is a holomorphic function independent of $\bar{\tau}$. More precisely, using modularity properties of the CFT together with spectral-flow invariance 
one can prove \cite{Kawai:1993jk} that the elliptic genus transforms as a weak Jacobi form~\cite{EZ} of index $m=c/6$ and weight 0.

In this work we will mainly be interested in the ellitpic genus of a CFT with target space\footnote{For some recent work on symmetry properties of such sigma models as well as the definition of twisted versions of $\phi$ (so called \emph{twining genera}) in these models see \cite{Gaberdiel:2011fg}.} $K3$ (with central charge $c=6$). In this case $\phi_{K3}$ is an universal quantity in the sense that it does not depend on the target space moduli and is given by the following rational function of Jacobi theta functions
\begin{align}
\phi_{K3}(\tau,z)&=8\left[\left(\frac{\theta_2(\tau,z)}{\theta_2(\tau,0)}\right)^2+\left(\frac{\theta_3(\tau,z)}{\theta_3(\tau,0)}\right)^2+\left(\frac{\theta_4(\tau,z)}{\theta_4(\tau,0)}\right)^2\right]\nonumber\\
&\simeq 2y+20+2y^{-1}+q\left(20y^2-128y+216-128y^{-1}+20y^{-2}\right)+\mathcal{O}(q^2)\,.\label{EllipticGenusDefinition}
\end{align}
In terms of the $\N=4$ algebra, $\phi_{K3}$ is sensitive to short and intermediate representations. Indeed, following \cite{Eguchi:1988vra,Eguchi:2009cq} (see also the more recent paper \cite{Eguchi:2010ej}) we can expand $\phi_{K3}$ 
\begin{align}
\phi_{K3}(\tau,z)&=24\ \text{ch}^{\cN=4}_{h=\tfrac{1}{4},\ell=0}(\tau,z)+\sum_{n=0}^\infty A_n\,\text{ch}^{\cN=4}_{h=n+\tfrac{1}{4},\ell=\tfrac{1}{2}}(\tau,z)\nonumber\\
&=24\ \text{ch}^{\cN=4}_{h=\tfrac{1}{4},\ell=0}(\tau,z)+\Sigma(\tau)\ \frac{\theta_1(\tau,z)^2}{\eta(\tau)^3}\ ,
\end{align}
where we have introduced (see \cite{Eguchi:1987sm})
\begin{align}
\text{ch}^{\cN=4}_{h,\ell=\tfrac{1}{2}}(\tau,z):=q^{h-\tfrac{3}{8}}\,\frac{\theta_1(\tau,z)^2}{\eta(\tau)^3}\,,&&\text{ch}^{\cN=4}_{h=\tfrac{1}{4},\ell=0}(\tau,z):=\frac{\theta_1(\tau,z)^2}{\eta(\tau)^3}\,\mu(\tau,z)
\end{align}
for the elliptic genera of the intermediate and short $\cN=4$ representations, respectively. The coefficients $A_n$ are positive integers and it is shown up to very high $n$ in \cite{Eguchi:2010ej,Cheng:2010pq,Gaberdiel:2010ch,Eguchi:2010fg}, that they can be decomposed into dimensions of irreducible representations of the Mathieu group $\mathbb{M}_{24}$ with non-negative integer multiplicities\footnote{This observation, which is known in the literature under the name \emph{Mathieu moonshine}, suggests the existence of a non-trivial action of $\mathbb{M}_{24}$ on the space of BPS states that contribute to $\phi_{K3}$.}. Furthermore, $\mu(\tau,z)$ is an Appell--Lerch sum defined as 
\begin{align}
\mu(\tau,z)=-\frac{ie^{\pi iz}}{\theta_1(\tau,z)}\sum_{n=-\infty}^\infty(-1)^n\frac{q^{\frac{n(n+1)}{2}}e^{2\pi inz}}{1-q^ne^{2\pi iz}}\ ,
\end{align}
and we also have used the weight $1/2$ mock modular form $\Sigma(\tau)$:
\be
\Sigma(\tau)=-8\left[\mu\left(\tau,u=\tfrac{1}{2}\right)+\mu\left(\tau,u=\tfrac{1+\tau}{2}\right)+\mu\left(\tau,u=\tfrac{\tau}{2}\right)\right]=-2\ q^{-\tfrac{1}{8}}\left(1-\sum_{n=1}^\infty A_nq^n\right)\,.
\ee
\section{BPS Amplitude and Short $\bm{\N=4}$ Representations}\label{Sect:12BPSandMock}
\setcounter{equation}{0}
In this section we consider a BPS--saturated one-loop amplitude with external legs stemming from $1/2$ BPS short multiplets of type II string theory compactified on $K3\times \T^2$. As we shall see, this amplitude  captures only the contribution of massless short multiplets to the elliptic genus of $K3$. As has already been discussed in the previous section (see also \cite{Eguchi:1988vra,Eguchi:2009cq}) this contribution can be written in terms of an Appell-Lerch sum.

\subsection{BPS--Saturated Amplitude}
\subsubsection{General Setup}
We start out by considering the following BPS--saturated one--loop coupling in type II string theory compactified on $K3\times \T^2$
\begin{align}
\Gc_{N}\,R_{(+),\mu\nu\rho\tau\sigma}R_{(+)}^{\mu\nu\rho\sigma}\, \left(F_{(-),\lambda\tau}F_{(-)}^{\lambda\tau}\right)^N\ \ \ ,\ \ \ N\in\mathbb{N}\ ,\label{EffectiveCouplingGauge}
\end{align}
with $R_{(+)}^{\mu\nu\rho\sigma}$ the self-dual part of the four-dimensional Riemann tensor and $F_{(-)}^{\mu\nu}$ the field strength of the Kaluza--Klein vector field stemming from the compactification on $\T^2$ (\emph{i.e.} the NS--NS graviphoton field strength tensor). We will choose all vertex operators to be inserted in the $(0,0)$--ghost picture and their generic expressions can be found in appendix~\ref{App:Vertex}. In order to simplify the computation, we  choose fixed helicities for all fields, i.e. to be concrete we evaluate the amplitude 
\begin{align}
&\amp_{N}(p_2,\bar{p}_2,\bar{p}_2^{(i)},p_2^{(i)})=\\
&\left\langle \int d^2z_1 V^{(0,0)}_R(h_{11},p_2)\int d^2z_2 V^{(0,0)}_R(h_{\bar{1}\bar{1}},\bar{p}_2)\, \prod_{i=1}^N\int d^2x_i V^{(0,0)}_{F}(\epsilon_1^{(i)},\bar{p}_2^{(i)})\int d^2y_iV^{(0,0)}_{F}(\epsilon_{\bar{1}}^{(i)},p_2^{(i)})\right\rangle\nonumber
\end{align}
In order to resemble the coupling (\ref{EffectiveCouplingGauge}) we need to extract the piece proportional to the momentum structure $p_2^2\bar{p}_2^2\prod_{i=1}^N\bar{p}_2^{(i)}p_2^{(i)}$. Comparing to (\ref{VertexGraviton}) and (\ref{VertexNSgauge}) we realize that both the bosonic as well as the fermion bilinear piece of the vertex operators will contribute. Therefore, the whole correlator can be written in the following way:
\begin{align}
\mathcal{G}_{N}&=\int d\zeta[N]\ \sum_{n=0}^N\big(^{2N}_{\,2n}\big)\left\langle\psi_1\psi_2(z_1)\,\bar{\psi}_1\bar{\psi}_2(z_2)\,\prod_{i=1}^n\psi_1\bar{\psi}_2(x_i)\,\bar{\psi}_1\psi_2(y_i)\right\rangle\ 
\left\langle\tilde{\psi}_1\tilde{\psi}_2(\bar{z}_1)\,\bar{\tilde{\psi}}_1\bar{\tilde{\psi}}_2(\bar{z}_2)\right\rangle\nonumber\\
&\times
\left\langle\prod_{j=n+1}^{N}(\bar{X}_2\partial X_1)(x_j)\,(X_2\partial\bar{X}_1)(y_j)\right\rangle\,\left\langle\prod_{i=1}^N\bar{\partial} X_3(\bar{x}_i)\,\bar{\partial} X_3(\bar{y}_i)\right\rangle 
=\sum_{n=0}^N\big(^{2N}_{\,2n}\big)\mathcal{G}_{N-n}^{\text{bos}}\, \mathcal{G}_{n}^{\text{ferm}}\ .
\label{CompleteAmplitude}
\end{align}
We have introduced the shorthand notation 
$\int d\zeta[N]:=\int d^2z_{1,2}\prod\limits_{i=1}^N\int d^2x_i\int d^2y_i$ for the integral measure.
In the following we compute the various correlators separately. 
\subsubsection{Bosonic Correlator}
The bosonic part of the correlator  (\ref{CompleteAmplitude}) takes the form
\begin{align}\label{35}
&\Gc_{N-n}^{\text{bos}}=(P_R)^{2N}\left\langle\prod_{j=n+1}^{N}\int d^2x_i(\bar{X}_2\partial X_1)(x_j)\int d^2y_i(X_2\partial\bar{X}_1)(y_j)\right\rangle\,,
\end{align}
where we have already used that the fields $\bar{\partial}X_3$ cannot contract with each other and therefore only contribute through their zero-modes. The correlator (\ref{35}) can be computed in a straight-forward way  using path integral methods. Indeed, following~\cite{Antoniadis:1995zn} we obtain:
\begin{align}
&\Gc_{N-n}^{\text{bos}}=(P_R)^{2N}\tau_2^{2N-2n}\left[\frac{\partial^{2N-2n}}{\partial w^{2N-2n}}G(w)\right]\bigg|_{w=0}\nonumber\\
&\equiv(P_R)^{2N}\tau_2^{2N-2n}\left[\frac{\partial^{2N-2n}}{\partial w^{2N-2n}}\sum_{g=1}^\infty\frac{w^{2g}}{(g!)^2\tau_2^{2g}}\,\left\langle\prod_{i=1}^g \int d^2 x_i\bar{X}_2\partial X_1 (x_i) \int d^2y_i X_2\bar{\partial}\bar{X}_1(y_i)\right\rangle\right]\bigg|_{w=0}\,.\nonumber
\end{align}
The generating functional $G(w)$ can be expressed in terms of the following normalized functional integral of two complex scalar fields
\be
G(w)=\frac{\int \prod_{i=1}^2\mathcal{D}X_i\mathcal{D}\bar{X}_i \text{Exp}\left(-S+\frac{w}{\tau_2}\int (\bar{X}_2\partial X_1+X_2\partial\bar{X}_1)\right)}{\int \prod_{i=1}^2\mathcal{D}X_i\mathcal{D}\bar{X}_i \text{Exp}\left(-S\right)}=
\left(\frac{2\pi iw\,\eta(\tau)^3}{\theta_1(w,\tau)}\right)^2e^{-\frac{2\pi w^2}{\tau_2}}\ ,\label{FunctionalIntegral}
\ee
where $S$ is the usual free-field action $S=\sum_{i=1,2}\frac{1}{\pi}\int d^2x(\partial X_i\bar{\partial}\bar{X}_i+\partial \bar{X}_i\bar{\partial}X_i)$. The evaluation of the functional integral has been performed in \cite{Antoniadis:1995zn} (see also\footnote{A particularly useful expansion of this expression in terms of standard Eisenstein series $E_{2k}(\tau)$ can e.g. be found in \cite{SW,Marino:1998pg} and is given by: 
$\frac{2\pi\eta^3 w}{\theta_1(w,\tau)}=-\text{Exp}\left\{\sum\limits_{k=1}^\infty\frac{\zeta(2k)}{k}\,E_{2k}(\tau)\,w^{2k}\right\}\ .$} \cite{Morales:1996bp}).
In total we find for the bosonic part of the amplitude:
\begin{align}
\Gc_{N-n}^{\text{bos}}&=(P_R)^{2N}\tau_2^{2N-2n}\left[\frac{\partial^{2N-2n}}{\partial w^{2N-2n}}\left(\frac{2\pi iw\,\eta(\tau)^3}{\theta_1(w,\tau)}\right)^2e^{-\frac{2\pi w^2}{\tau_2}}\right]\bigg|_{w=0}\,.\label{BosonicCorrAmp}
\end{align}
\subsubsection{Fermionic Correlator}\label{Sect:FermCorr}
To compute the fermionic part of the correlator (\ref{CompleteAmplitude}) we work at a generic point in the moduli space of $K3$ and use the following table summarizing the charges in the $SO(2)\times \Gamma$ lattice (see appendix \ref{App:GenK3Compact} for more details): 
\begin{center}
\begin{tabular}{|c||c||c||c|c|c|c||c|c|c|c|}\hline
&&&&&&&&&&\\[-10pt]
\textbf{vertex} & \# & \textbf{pos.} & $\phi_1$ & $\phi_2$ & $H_3$ & $H$ & $\tilde{\phi}_1$ & $\tilde{\phi}_2$ & $\tilde{H}_3$ & $\tilde{H}$ \\\hline\hline
&&&&&&&&&&\\[-10pt]
graviton & $1$ & $z_1$ & $+1$ & $+1$ & $0$ & $0$ & $+1$ & $+1$ & $0$ & $0$ \\[2pt]\hline
&&&&&&&&&&\\[-10pt]
& $1$ & $z_2$ & $-1$ & $-1$ & $0$ & $0$ & $-1$ & $-1$ & $0$ & $0$ \\[2pt]\hline\hline
&&&&&&&&&&\\[-10pt]
vector field & $n$ & $x_i$ & $+1$ & $-1$ & $0$ & $0$ & $0$ & $0$ & $0$ & $0$ \\[4pt]\hline
&&&&&&&&&&\\[-10pt]
& $n$ & $y_i$ & $-1$ & $+1$ & $0$ & $0$ & $0$ & $0$ & $0$ & $0$ \\[4pt]\hline
\end{tabular}
\end{center}
With this information it is straightforward to write down the fermionic contribution
\begin{align}
\Gc_n^{\text{ferm}}:&=\int d\zeta[n]\ \left\langle\psi_1\psi_2(z_1)\,\bar{\psi}_1\bar{\psi}_2(z_2)\,\prod_{i=1}^n\psi_1\bar{\psi}_2(x_i)\,\bar{\psi}_1\psi_2(y_i)\right\rangle\left\langle\tilde{\psi}_1\tilde{\psi}_2(\bar{z}_1)\,\bar{\tilde{\psi}}_1\bar{\tilde{\psi}}_2(\bar{z}_2)\right\rangle\nonumber\\
&=\int d\zeta[n]\ \sum_s\frac{F_{\Lambda,s}\left(z_1-z_2+\sum\limits_{i=1}^n(x_i-y_i),z_1-z_2-\sum\limits_{i=1}^n(x_i-y_i),0\right)}{E^2(z_1,z_2)\prod_{i,j}^nE^2(x_i,y_i)} \nonumber\\
&\hspace{1.7cm}\times\sum_{\tilde{s}}\frac{\tilde{F}_{\Lambda,\tilde{s}}\left(\bar{z}_1-\bar{z}_2,\bar{z}_1-\bar{z}_2,0\right)}{E^2(\bar{z}_1,\bar{z}_2)}\ ,
\end{align}
where we have already used that the ghost contribution has canceled a $\vartheta$-function corresponding to the $\phi_3$-plane. Here $\sum_{s,\tilde{s}}$ denotes the sum over all possible spin structures, which can be explicitly performed using (\ref{GenExprSpinStruct}): 
\begin{align}
&\Gc_n^{\text{ferm}}=\int d\zeta[n]\ \frac{F_\Lambda\left(z_1-z_2,z_1-z_2,\sqrt{2}\sum\limits_{i=1}^n(x_i-y_i)\right)}{E^2(z_1,z_2)\prod_{i,j}^nE^2(x_i,y_i)}\ \frac{\tilde{F}_\Lambda\left(\bar{z}_1-\bar{z}_2,\bar{z}_1-\bar{z}_2,0\right)}{E^2(\bar{z}_1,\bar{z}_2)}\ .\nonumber
\end{align}
Using (\ref{FLtheta}) together with the explicit definition of the prime forms $E(z_1,z_2)$ this expression can be rewritten in the following manner
\begin{align}
&\Gc_n^{\text{ferm}}=\int d\zeta[n]\ \frac{\Theta_\Lambda\left(\sqrt{2}\sum\limits_{i=1}^n(x_i-y_i)\right)}{\prod_{i,j}^nE^2(x_i,y_i)}\ .
\end{align}
Using the bosonization identities developed in \cite{Verlinde:1986kw} we can rewrite this expression in terms of a correlator of the internal 2-dimensional CFT with target space $K3$
\begin{align}
&\Gc_n^{\text{ferm}}=4\tau_2\,\left\langle\prod_{i=1}^n\int d^2x_ie^{\sqrt{2}i H}(x_i)\,\int d^2y_ie^{-\sqrt{2}iH}(y_i)\right\rangle_{K3}\,.\label{K3corrEllGen}
\end{align}
This expression is however just the $2n$-th derivative of the elliptic genus of $K3$, i.e.
\begin{align}
\Gc_n^{\text{ferm}}&=4\tau_2\left[\frac{\partial^{2n}}{\partial z^{2n}}\phi_{K3}(\tau,z)\right]\bigg|_{z=0}\,,
\end{align}
which has been introduced in (\ref{EllGenDef}). The result can either be directly obtained by rewriting the insertions in (\ref{K3corrEllGen}) in terms of the neutral currents $J_{K3}$ or by direct computation of the world-sheet $x_i$ and $y_i$ integrations in (\ref{K3corrEllGen}). The latter are most easily tackled by working at a particular point in the $K3$-moduli space. 
In fact, we perform\footnote{However, the calculations can 
easily be generalized to other orbifold points in the $K3$ moduli space along~\cite{Stieberger:1998yi}.} the calculation in an $\mathbb{Z}_2$  orbifold limit of K3. 
In this case (\ref{K3corrEllGen}) can be written as the following  correlator of fermion bilinear insertions:
\begin{align}
\Gc_n^{\text{ferm}}=4\tau_2\,\sum_{h,g=0,\tfrac{1}{2}}\left\langle\prod_{i=1}^n\int d^2x_i\psi_4\psi_5(x_i)\,\int d^2y_i\bar{\psi}_4\bar{\psi}_5(y_i)\right\rangle_{(h,g)}\,.\label{AmpEllGenRes}
\end{align}
Correlators of this type have already been considered in \cite{Stieberger:2002wk} by breaking them down into smaller individual contractions. Indeed, denoting the fermion one-loop propagator for a particular even spin-structure $\vec{\beta}=(\beta_1,\beta_2)$ by
\begin{align}
G^F_{\vec{\beta}}(x-y):=\langle\psi_4(x)\bar{\psi}_4(y)\rangle_{\vec{\beta}}=\langle\psi_5(x)\bar{\psi}_5(y)\rangle_{\vec{\beta}}=\frac{2\pi \eta(\tau)^3\theta\big[^{\beta_1}_{\beta_2}\big](\tau,x-y)}{\theta\big[^{1/2}_{1/2}\big](\tau,x-y)\theta\big[^{\beta_1}_{\beta_2}\big](\tau,0)}\,,
\end{align}
the following equality is proven in \cite{Stieberger:2002wk} for $M$ even and $M>2$
\begin{align}
&\prod_{a=1}^M\int d^2x_a\, G^F_{\vec{\beta}}(x_1-x_2)\,G^F_{\vec{\beta}}(x_2-x_3)\ldots G^F_{\vec{\beta}}(x_M-x_1)
=-\frac{(2\tau_2)^M}{(M-1)!}\left[\frac{\partial^M}{\partial z^M}\,\ln\theta_{\vec{\beta}}(z,\tau)\right]\bigg|_{z=0}\nonumber\\
&\hskip3cm=2\zeta(M)(2\tau_2)^M\left[2^{2\beta_1 M}E_M(4^{2\beta_1}\tau/2+\beta_1+\beta_2+1/2)-E_M(\tau)\right]\,,
\end{align}
where $E_{M}$ denotes the $M$--th Eisenstein series. For $M=2$ a similar expression holds with an additional non-holomorphic shift term:
\begin{align}
\int d^2x_1\int d^2x_2\, G^F_{\vec{\beta}}(x_1-x_2)^2=(2\tau_2)^2\left(\left[\frac{\partial^2}{\partial z^2}\,\ln\theta_{\vec{\beta}}(z,\tau)\right]\bigg|_{z=0}+\frac{\pi}{\tau_2}\right)\,.
\end{align}
Indeed, using these expressions, we can assemble the full amplitude (\ref{AmpEllGenRes}) by summing over all  contractions with the appropriate normalization factors, which can  directly be taken over from~\cite{Lerche:1999ju}. For convenience we  compile the first few examples in table 1 and obtain for the fermion correlator $g_n(\tau)$ the final result:
\begin{align}
g_2(\tau)&=E_2 \nonumber\\
g_4(\tau)&=E_4 +E_2^2 \ ,\nonumber\\
g_6(\tau)&=-\frac{1}{12}\ \left(\ 8E_6 -15E_4 E_2 -5E_2^3 \ \right)\ ,\nonumber\\
g_8(\tau)&=\frac{1}{72}\ \left[\ 210E_4 E_2^2 -224E_6 E_2 +51E_4^2 +35E_2^4 \ \right]\ ,\nonumber\\
g_{10}(\tau)&=\frac{1}{48}\big[\ 255E_4^2 E_2 +350E_4 E_2^3 -560E_6 E_2^2 
-32E_6 E_4 +35E_2^5\ \big]\ ,\nonumber\\
g_{12}(\tau)&=\frac{1}{288}\big[\ 256E_6 ^2-2112E_6 E_4 E_2 -111E_4^3-12320E_6 E_2^3 +8415E_4^2 E_2^2\nonumber\\
&\hskip1.25cm +5775E_4 E_2^4 +385E_2^6 \big]\ .
\end{align}

\parbox{17.5cm}{\begin{minipage}[t]{.45\textwidth}\vspace{0pt}
\begin{tabular}{|c|r|r|}\hline
$n$ & \textbf{contr.} & \textbf{mult.} \\\hline
&&\\[-8pt]
$2$ & $[2]$ & $1$ \\[4pt]\hline
&&\\[-8pt]
$4$ & \parbox{0.9cm}{$\hspace{0.4cm}[4]$ \\ $[2,2]$} & \parbox{0.5cm}{$\phantom{-}2$ \\ $-2$} \\[10pt]\hline
&&\\[-8pt]
$6$ & \parbox{1.21cm}{$\hspace{0.79cm}[6]$ \\ ${}\hspace{0.39cm}[2,4]$ \\ $[2,2,2]$} & \parbox{0.7cm}{$-12$ \\ $\phantom{-}18$ \\ \phantom{}$\hspace{0.2cm}-6$} \\[18pt]\hline
&&\\[-8pt]
$8$ & \parbox{1.6cm}{${}\hspace{1.17cm}[8]$ \\ ${}\hspace{0.39cm}[2,2,4]$ \\ ${}\hspace{0.79cm}[2,6]$ \\ ${}\hspace{0.79cm}[4,4]$ \\ $[2,2,2,2]$} & \parbox{0.9cm}{$-144$\\$-144$ \\ $\phantom{-}192$ \\ ${}\hspace{0.52cm}72$ \\ ${\hspace{0.52cm}}24$} \\[32pt]\hline 
\end{tabular}
\end{minipage}
\hspace{-3cm}
\begin{minipage}[t]{.45\textwidth}\vspace{0pt}
\begin{tabular}{|c|c|c|}\hline
$n$ & \textbf{contr.} & \textbf{mult.} \\\hline
&&\\[-8pt]
$10$ & \parbox{2cm}{${}\hspace{1.37cm}[10]$ \\ ${}\hspace{1.19cm}[2,8]$ \\ ${}\hspace{1.19cm}[4,6]$ \\ ${}\hspace{0.79cm}[2,2,6]$ \\ ${}\hspace{0.79cm}[2,4,4]$ \\ ${}\hspace{0.39cm}[2,2,2,4]$ \\ $[2,2,2,2,2]$} & \parbox{1.15cm}{$-2880$\\$\phantom{-}3600$ \\ $\phantom{-}2400$ \\ $-2400$ \\ $-1800$ \\ $\phantom{-}1200$ \\ ${}\hspace{0.1cm}-120$}\\[47pt]\hline 
\end{tabular}
\end{minipage}
\hspace{-2.3cm}
\begin{minipage}[t]{.45\textwidth}\vspace{0pt}
\begin{tabular}{|c|c|c|}\hline
$n$ & \textbf{contr.} & \textbf{mult.} \\\hline
&&\\[-8pt]
$12$ & \parbox{2.4cm}{${}\hspace{1.77cm}[12]$ \\ ${}\hspace{1.38cm}[2,10]$ \\ ${}\hspace{1.59cm}[4,8]$ \\ ${}\hspace{1.59cm}[6,6]$ \\ ${}\hspace{1.19cm}[2,2,8]$ \\ ${}\hspace{1.19cm}[2,4,6]$ \\${}\hspace{1.19cm}[4,4,4]$ \\ ${}\hspace{0.79cm}[2,2,2,6]$ \\ ${}\hspace{0.79cm}[2,2,4,4]$ \\ ${}\hspace{0.39cm}[2,2,2,2,4]$ \\ $[2,2,2,2,2,2]$} & \parbox{1.35cm}{$-86400$\\${}\hspace{0.12cm}103680$ \\ $\phantom{-}64800$ \\ $\phantom{-}28800$ \\ $-64800$ \\ $-86400$ \\ $-10800$ \\ $\phantom{-}28800$ \\ $\phantom{-}32400$ \\ $-10800$ \\ ${}\hspace{0.74cm}720$} \\[75pt]\hline 
\end{tabular}
\end{minipage}

\noindent
\begin{center}
\hspace{-2cm}\parbox{15.5cm}{Table 1: {\it Overview over the cyclic contractions in the fermion correlator (\ref{AmpEllGenRes}) for even $n=2,4,6,8,10,12$ (the correlator is identically zero for $n$ odd) with fixed ordering of the vertices. The column 'contr.' indicates how the original correlation function in (\ref{AmpEllGenRes}) is decomposed into shorter correlators. Indeed, the notation $[m_1,\ldots,m_k]$ means that (\ref{AmpEllGenRes}) has been broken into $k$ correlators involving $m_i$ fermion pairs respectively (for more details on this notation see  \cite{Lerche:1999ju}). The third column displays the multiplicity of each decomposition. }}
\end{center}}
${}$\\

\noindent
Notice that the multiplicities of the various types of contractions, displayed in the third  column, give the relative multiplicities for a given particular ordering of the vertices. Indeed, to obtain the correct result, we still need to add an overall factor of $\frac{2^{2n}\tau_2^{2n}(n!)^2}{(2n)!}$ at each step, which takes into account the appropriate combinatorial factors.
Taking all contributions together, we indeed find:
\begin{align}
\Gc^{\text{ferm}}_n&=
4\tau_2^{2n+1}\left[\frac{\partial^{2n}}{\partial z^{2n}}\phi_{K3}(\tau,z)\right]\bigg|_{z=0}\,.
\end{align}

\subsubsection{Complete Coupling and Holomorphic Limit}\label{Sect:MockModularForm}
Having worked out the bosonic and fermionic part of the amplitude we are in a position to put them together and assemble the full amplitude
\begin{align}
\mathcal{G}_N=4\tau_2^{2N+1}(P_R)^{2N}\sum_{n=0}^N\big(^{2N}_{\,2n}\big)\frac{\partial^{2n}}{\partial z^{2n}}\frac{\partial^{2N-2n}}{\partial w^{2N-2n}}\left(\frac{2\pi iw\,\eta(\tau)^3}{\theta_1(w,\tau)}\right)^2\,\phi_{K3}(\tau,z)e^{-\frac{2\pi (w^2+z^2)}{\tau_2}}\bigg|_{z=w=0}\,.
\end{align}
We can explicitly perform the sum over $n$ yielding the result
\begin{align}
\mathcal{G}_N=-16\pi^2\tau_2^{2N+1}(P_R)^{2N}\left(\frac{\partial}{\partial z}+\frac{\partial}{\partial w}\right)^{2N}\left[\frac{w^2\,\eta(\tau)^6}{\theta_1(w,\tau)^2}\,\phi_{K3}(\tau,z)\,e^{-\frac{2\pi (w^2+z^2)}{\tau_2}}\,\right]\bigg|_{z=w=0}\,.
\end{align}
To further simplify this expression, we introduce the new variables
\begin{align}
&u=\frac{1}{2}(z+w)\,,&&\text{and} &&v=\frac{1}{2}(z-w)\,,
\end{align}
such that the amplitude becomes
\begin{align}
\mathcal{G}_N&=-16\pi^2\tau_2^{2N+1}(P_R)^{2N}\frac{\partial^{2N}}{\partial u^{2N}}\left[\frac{(u-v)^2\,\eta(\tau)^6}{\theta_1(u-v,\tau)^2}\,\phi_{K3}(\tau,u+v)\,e^{-\frac{4\pi (u^2+v^2)}{\tau_2}}\,\right]\bigg|_{u=v=0}=\nonumber\\
&=-16\pi^2\tau_2^{2N+1}(P_R)^{2N}\frac{\partial^{2N}}{\partial u^{2N}}\left[\frac{u^2\,\eta(\tau)^6}{\theta_1(u,\tau)^2}\,\phi_{K3}(\tau,u)\,e^{-\frac{4\pi u^2}{\tau_2}}\,\right]\bigg|_{u=0}\,.\label{FinalAmp}
\end{align}
Note that the expression (\ref{FinalAmp}) is strongly related to the elliptic genus of $K3$. In fact, in the holomorphic limit, it just corresponds to the contribution of the short multiplets with $\ell=0$. To see this, we will for the moment ignore the factor $e^{-\frac{4\pi u^2}{\tau_2}}$ and moreover recall from section~\ref{App:EllGenK3} that the elliptic genus can be written in the following manner
\begin{align}
\phi_{K3}(\tau,u)=\frac{\theta_1(\tau,u)^2}{\eta(\tau)^3}\ \left[\ 24\ \mu(\tau,z)+\Sigma(\tau)\ \right]\ ,
\end{align}
where the object $\mu(\tau,u)$ is a mock theta function, which has attracted a lot of attention recently both in mathematics and physics (see \cite{Zagier2} and references therein for a very nice overview). With this we obtain for (the holomorphic limit of) (\ref{FinalAmp})
\begin{align}
\mathcal{G}^{\text{hol}}_N=\left\{\begin{array}{lcl}-16\pi^2\tau_2^3(P_R)^2\eta(\tau)^3\left[\ 24\tfrac{\partial^2}{\partial u^2}\left(u^2\mu(\tau,u)\right)\big|_{u=0}+\Sigma(\tau)\ \right] &, & N=1 \\&&\\-384\pi^2\tau_2^{2N+1}(P_R)^{2N}\eta(\tau)^3\tfrac{\partial^N}{\partial u^N}\left(u^2\mu(\tau,u)\right)\big|_{u=0} &, & N>1\end{array}\right.\label{N2partAmp}
\end{align}
Notice that the case $N=1$ is somewhat different than the remaining series. We will briefly return to this point in section~\ref{Sect:N1Amp} where we will reinterpret this expression by calculating a slightly different (but supersymmetrically related) amplitude.
\subsection[Holomorphic Anomaly and $1/2$ BPS Contribution]{Holomorphic Anomaly and $\bm{1/2}$ BPS Contribution}\label{Sect:HolAnom}
To better understand the holomorphic limit, which we have discussed in the previous subsection, 
we shall now consider the analog of the holomorphic anomaly equation \cite{Bershadsky:1993cx,Antoniadis:1993ze} for the coupling (\ref{FinalAmp}). To this end, let us also explicitly include the integration over the world-sheet torus as well as the Siegel--Narain theta-function for the torus compactification:
\begin{align}
\mathcal{G}_N&=-16\pi^2\int \frac{d^2\tau}{\tau_2}\tau_2^{2N}\frac{\partial^{2N}}{\partial u^{2N}}\left[\frac{u^2\,\eta(\tau)^6}{\theta_1(u,\tau)^2}\,\phi_{K3}(\tau,u)\,e^{-\frac{4\pi u^2}{\tau_2}}\,\right]\bigg|_{u=0}\nonumber \\[3mm]
&\times\sum_{(P_L,P_R)\in\Gamma^{2,2}} (P_R)^{2N}\,q^{\frac{1}{2}|P_L|^2}\bar{q}^{\frac{1}{2}|P_R|^2}\ .
\end{align}
As we shall discuss at length in the next section, the integral $\mathcal{G}_N$ does not arbitrarily depend on the VM moduli, but is rather a function of just a particular harmonic-projection of the $\N=4$ VMs. For example (following a similar reasoning as in \cite{Antoniadis:1993ze}) one would (for our conventions of Narain momenta (\ref{NarainMomenta})) expect $\mathcal{G}_N$ to be independent of the modulus $T$. To check this we apply a (covariant) $T$-derivative in the following manner
\be
\mathcal{D}_T\mathcal{G}_N:=32\pi^2iT_2\left(\frac{\partial}{\partial T} +\frac{iN}{4T_2}\right)\int \frac{d^2\tau}{\tau_2}\tau_2^{2N}\,\mathcal{P}_{2N}(\tau,\bar{\tau})\,\sum_{(P_L,P_R)\in\Gamma^{2,2}} (P_R)^{2N}\,q^{\frac{1}{2}|P_L|^2}\bar{q}^{\frac{1}{2}|P_R|^2}\ ,\label{HolAnomEqu}
\ee
where we have introduced the weight $(2N,0)$ modular form:
\begin{align}
\mathcal{P}_{2N}(\tau,\bar{\tau}):=\frac{\partial^{2N}}{\partial u^{2N}}\left[\frac{u^2\,\eta(\tau)^6}{\theta_1(u,\tau)^2}\,\phi_{K3}(\tau,u)\,e^{-\frac{4\pi u^2}{\tau_2}}\,\right]\bigg|_{u=0}\,.\label{GenFunctInteg}
\end{align}
We stress that $\mathcal{P}_{2N}(u,\tau,\bar{\tau})$ is non-holomorphic only because of the presence of $e^{-4\pi u^2/\tau_2}$ and satisfies the recursive relation
\begin{align}
\frac{\partial}{\partial \bar{\tau}}\mathcal{P}_{2N}(\tau,\bar{\tau})=\frac{2\pi i}{\tau_2^2}\,\mathcal{P}_{2N-2}(\tau,\bar{\tau})\,.\label{RecursionTau}
\end{align}
Evaluating (\ref{HolAnomEqu}) explicitly we find
\begin{align}
\mathcal{D}_T\mathcal{G}_N&=-16\pi^2\int d^2\tau\mathcal{P}_{2N}(\tau,\bar{\tau})\,\sum_{(P_L,P_R)} P_L (P_R)^{2N-1}\left[2N-2\pi \tau_2 |P_R|^2\right]\tau_2^{2N-1}\,q^{\frac{1}{2}|P_L|^2}\bar{q}^{\frac{1}{2}|P_R|^2}\nonumber\\
&=-16\pi^2\int d^2\tau\mathcal{P}_{2N}(\tau,\bar{\tau})\,\frac{\partial}{\partial\bar{\tau}}\sum_{(P_L,P_R)} P_L (P_R)^{2N-1}\,\tau_2^{2N}\,q^{\frac{1}{2}|P_L|^2}\bar{q}^{\frac{1}{2}|P_R|^2}\ .
\end{align}
Performing a partial integration in $\bar{\tau}$ does not give rise to a bounday contribution for $N>1$. 
Hence by using (\ref{RecursionTau}) we find
\begin{align}
\mathcal{D}_T\mathcal{G}_N&=-32\pi^3i\int \frac{d^2\tau}{\tau_2}\tau_2^{2N-1}\,\mathcal{P}_{2N-2}(\tau,\bar{\tau})\sum_{(P_L,P_R)} P_L (P_R)^{2N-1}\,q^{\frac{1}{2}|P_L|^2}\bar{q}^{\frac{1}{2}|P_R|^2}\nonumber\\
&=32\pi^2U_2\left(\frac{\partial}{\partial \bar{U}}+\frac{i(N-1)}{4U_2}\right)\int \frac{d^2\tau}{\tau_2}\tau_2^{2N-2}\,\mathcal{P}_{2N-2}(\tau,\bar{\tau})\sum_{(P_L,P_R)} (P_R)^{2N-2}\,q^{\frac{1}{2}|P_L|^2}\bar{q}^{\frac{1}{2}|P_R|^2}\nonumber\\
&=:32\pi^2\mathcal{D}_{\bar{U}} \mathcal{G}_{N-1}\,,
\end{align}
where we have introduced the covariant derivative with respect to $\bar{U}$. To summarize, we have found the following recursive relation
\begin{align}
\mathcal{D}_T\mathcal{G}_N=32\pi^2\mathcal{D}_{\bar{U}} \mathcal{G}_{N-1}\,, 
\end{align}
where the right hand side depends on the same $\mathcal{G}$, however, with reduced index. Following the same spirit as \cite{Antoniadis:1993ze}, we therefore call this contribution 'anomalous'. In the next section
we discuss its origin from an effective action point of view. However, here we notice that from the point of view of the amplitude, the anomaly  is due to the $e^{-4\pi u^2/\tau_2}$ factor in the generating functional (\ref{GenFunctInteg}). Thus taking the holomorphic limit (as we have considered in the previous subsection) 
removes the anomaly and  provides the 'purely topological' coupling, which in this case corresponds to the 
$1/2$ BPS contribution of short multiplets to the elliptic genus of $K3$.
\section{Superspace Analysis and Effective Action Couplings}\label{Sect:SuperspaceGrav}
\setcounter{equation}{0}
Before continuing, we  investigate the amplitudes discussed in the previous section from the point of view of the effective action to understand their BPS structure. This is best illuminated by formulating these couplings in superspace, which is suitable for manifestly displaying invariance under supersymmetry. Four dimensional theories with $\cN=4$ supersymmetry have 16 supercharges and can therefore be best described in terms of a $(4|16)$ dimensional superspace. Instead of using the standard $\mathbb{R}^{4|16}$, we use harmonic superspace, which will turn out to be better suited for the description of our couplings (for a review of the relevant conventions and notation see appendix~\ref{App:HarmSuperspace}). 

The spectrum of type II string theory in $\cN=4$ compactifications contains two types of BPS states: $1/2$  and $1/4$ BPS states, which group themselves in short and intermediate multiplets, respectively. They are subject to particular analyticity properties such that,  roughly speaking, short multiplets depend only on half of the superspace coordinates. It turns out, that the couplings discussed in the previous section can be entirely described using short multiplets, which we have reviewed for the reader's convenience in appendix~\ref{App:LinOnShellMultiplets}. 

Before discussing the actual couplings we should also note that for the sake of manifest $SO(6,22)$ covariance, we will formulate all couplings in the 'supergravity frame'. Indeed, as already discussed in section~\ref{SPEC}, in $\N=4$ supergravity, the $SO(6,22)$ symmetry is linearized by introducing six additional \emph{compensator} VMs (see \cite{Antoniadis:2007cw,Antoniadis:2009tr} for a more detailed discussion). In order to reduce to only physical fields, there are  two possibilities: Either the gauge fields of the compensating multiplets are expressed as functions of the graviphotons which sit inside the supergravity multiplet ('superstring basis') or the relation is inverted and the graviphotons are identified with the gauge fields of the compensating multiplets; in this case, the vector bosons of the supergravity multiplet are expressed as functions of all VM gauge fields ('supergravity basis'). While for explicit amplitude computations in superstring theory, the former basis is relevant,  in this section, to display manifest $SO(6,22)$ invariance,  we stick to the latter frame.

\subsection[Leading On-shell $1/4$ BPS Protected Coupling]{Leading On-shell $\bm{1/4}$ BPS Protected Coupling}\label{Sect:N1Amp}

Couplings involving $1/2$ BPS short multiplets (Weyl and VMs) have already been discussed earlier in \cite{Antoniadis:2006mr,Antoniadis:2007cw}. For concreteness, here we will start out by considering a particular harmonic projection of the VM (\ref{N4shortVectormulti}) (see also~\cite{Andrianopoli:1999vr}), {\it e.g.}
\begin{align}
&Y^{12}_A=Y^{12}_A(z,\theta_3,\theta_4,\bar{\theta}^1,\bar{\theta}^2,u)\,,&&\text{with} &&{z^{\alpha}}_{\dot{\beta}}=x_\mu{(\sigma^\mu)^\alpha}_{\dot{\beta}}+i\left(\theta_3^\alpha\bar{\theta}^3_{\dot{\beta}}+\theta_4^\alpha\bar{\theta}^4_{\dot{\beta}}-\theta_1^\alpha\bar{\theta}^1_{\dot{\beta}}-\theta_2^\alpha\bar{\theta}^2_{\dot{\beta}}\right)\,,\label{ProjVectMult1}
\end{align}
where we have also added an index of the $SO(6,22)$ VM gauge group. This multiplet satisfies the analyticity properties
\begin{align}
D^1_\alpha Y^{12}_A=D^2_\alpha Y^{12}_A=\bar{D}_3^{\dot{\alpha}} Y^{12}_A=\bar{D}_4^{\dot{\alpha}} Y^{12}_A=0\,.
\end{align}
Thus, this multiplet can be consistently coupled to the Weyl multiplet in the following form
\begin{align}
S&=\int d^4x\int du\int d^2\theta_1\int d^2\theta_2\int d^2\theta_3\int d^2\theta_4\int d^2\bar{\theta}^1\int d^2\bar{\theta}^2\, \mathbb{G}(W,Y_A^{12},u)\nonumber\\
&=\int d^4x\int du\int d^2\theta_3\int d^2\theta_4\int d^2\bar{\theta}^1\int d^2\bar{\theta}^2\, (D^1\cdot D^1)(D^2\cdot D^2)\left[\mathbb{G}(W,Y_A^{12},u)\right]\,,\label{S14GravGen}
\end{align}
for some coupling function $\mathbb{G}$ depending also on the Weyl multiplet. This expression is an integral over $12$ Grassmann coordinates and for the purpose of calculating explicit string theory amplitudes we are interested in the component expansion. Indeed, the four spinor derivatives we have explicitly written out can only hit the Weyl multiplets in the above coupling. Performing also the Grassmann integrals, we will find among others the following term at the component level
\begin{align}
S=\int d^4x\int du &\left(R^{(+),\mu\nu\rho\tau}R^{(+)}_{\mu\nu\rho\tau}\right)\bigg[\left(F_{(-)}^{A}\cdot F_{(-)}^{B}\right)+F_{(-),\lambda\sigma}^{A}\left(\bar{\lambda}^1_{B}\sigma^{\lambda\sigma}\bar{\lambda}^2_{C}\right)\frac{\partial}{\partial\varphi^{12}_{C}}\bigg]\mathcal{G}_{AB}(\Phi,\varphi^{12}_A,u)\,,\label{Grav14Comp}
\end{align}
where we have introduced the shorthand notation
\begin{align}
\mathcal{G}_{AB}(\Phi,\varphi^{12}_A,u)=\frac{\partial^4\mathbb{G}(W,Y_A^{12},u)}{\partial W^2\partial Y^{12}_{A}\partial Y^{12}_{B}}\bigg|_{\theta=0}\,.
\end{align}
The second term in the square brackets of (\ref{Grav14Comp}) has just been added to show that there are also further component couplings leading to the same $\mathcal{G}_{AB}$. Indeed, in order to check this, we have computed the coupling 
\begin{align}
\tilde{\mathcal{G}}_{A_1A_2A_3}\left(R^{(+)}_{\mu\nu\rho\tau}R^{(+),\mu\nu\rho\tau}\right)\, F^{(-),A_1}_{\lambda\sigma}\left(\bar{\lambda}_1^{A_2}\bar{\sigma}^{\lambda\sigma}\bar{\lambda}_2^{A_3}\right)\,,
\end{align}
where $\lambda_{1,2}$ denote T-modulini (\emph{i.e.} superpartners of the Kaluza-Klein vector fields) and $F^{(-)}$ is the field strength of a Kaluza-Klein vector field. Through a straight-forward computation one can show
\begin{align}
&\tilde{\mathcal{G}}_{A_1A_2A_3}=\mathcal{D}_{A_2}^{12}\mathcal{G}_{A_1A_3}\,,&&\text{with} && \mathcal{G}_{A_1A_3}=\int_{\mathcal{M}}\left\langle\int d^2w\,J_{K3}\hat{V}_{A_1}(w)\int d^2y\,J_{K3}\hat{V}_{A_3}(y)\right\rangle+\mathcal{C}\,,\nonumber
\end{align}
where $\mathcal{C}$ is an arbitrary function which is independent of the (massless) VM moduli $\varphi_A^{12}$ which we will drop in the following. Notice that this is exactly the relation we have anticipated based on the superspace coupling (\ref{Grav14Comp}). Choosing now a setup in which the $\hat{V}_{A_1,A_3}$ only contribute the bosonic zero modes on the $\T^2$ and dropping a constant tensor which takes care of the $SO(n)$-structure, we can rewrite this expression as:
\begin{align}
\mathcal{G}\sim\text{Tr}_{RR}\left[(-1)^{F+\bar{F}}(J_0)^2q^{L_0-\frac{1}{4}}\bar{q}^{\bar{L}_0-\frac{1}{4}}\right]\ (P_R)^2\ .
\end{align}
Combining this result with its conjugate, we obtain
\begin{align}
\mathcal{G}+\bar{\mathcal{G}}&\sim\text{Tr}_{RR}\left\{(-1)^{F+\bar{F}}\left[(J_0P_R)^2+(\bar{J}_0P_L)^2\right]q^{L_0-\frac{1}{4}}\bar{q}^{\bar{L}_0-\frac{1}{4}}\right\}\nonumber\\
&=-\frac{1}{4\pi^2}\left(\frac{\partial}{\partial z}+\frac{\partial}{\partial\bar{z}}\right)^2\text{Tr}_{RR}\left[(-1)^{F+\bar{F}} e^{2\pi i z J_0 P_R} e^{-2\pi i \bar{z} \bar{J}_0 P_L}q^{L_0-\frac{1}{4}}\bar{q}^{\bar{L}_0-\frac{1}{4}}\right]\bigg|_{z=\bar{z}=0}\nonumber\\
&=-\frac{1}{4\pi^2}\left(\frac{\partial}{\partial z}+\frac{\partial}{\partial\bar{z}}\right)^2 Z(z P_R,\bar{z}P_L)\bigg|_{z=\bar{z}=0}\,,\label{DefHelTr6}
\end{align}
where $Z(z,\bar{z})$ is the generating functional of the helicity supertraces as defined in (\ref{GenFuncHelTr}). Expression (\ref{DefHelTr6}) resembles very closely the definition (\ref{HelStrDef}) of the helicity supertrace $B_6$, which is indeed an index of the $1/2$ BPS short multiplets only.
\subsection[Higher Point $1/4$ BPS Protected Couplings]{Higher Point $\bm{1/4}$ BPS Protected Couplings}\label{Sect:K3genSeries}
The strategy in finding superspace couplings also for higher $N$ is to consider a particular superdescendant of the VM
\begin{align}
\Upsilon_A^{\mu\nu}:=(D^3\sigma^{\mu\nu}D^4)Y^{12}_A\,,
\end{align}
whose lowest component is indeed $F_{(-),A}$. With this we can then consider the extended coupling
\begin{align}
S_N&=\int d^4x\int du\int d^2\theta_{3,4}\int d^2\bar{\theta}^{1,2}(D^1\cdot D^1)(D^2\cdot D^2)\left[\prod_{i=1}^{N-1}(\Upsilon_{A_i}\cdot \Upsilon_{B_i})\mathbb{G}^{A_iB_i}(W,Y_A^{12},u)\right].\label{S14GravGenEx}
\end{align}
As before, evaluating all spinor derivatives and performing explicitly all the Grassmann integrations, we find (among others) the following term at the component level
\begin{align}
S_N&=\int d^4x\int du\left(R^{(+),\mu\nu\rho\tau}R^{(+)}_{\mu\nu\rho\tau}\right)\prod_{i=1}^N\left(F_{(-)}^{A}\cdot F_{(-)}^{B}\right) \mathcal{G}_{A_1\ldots A_NB_1\ldots B_N}(\Phi,\varphi^{12}_A,u)+\ldots\,,
\end{align}
where we have introduced the coupling functions
\begin{align}
\mathcal{G}_{A_1\ldots A_NB_1\ldots B_N}(\Phi,\varphi^{12}_A,u)=\frac{\partial^4\mathbb{G}_{A_2\ldots A_NB_2\ldots B_N}(W,Y_{A}^{12},u)}{\partial W^2\partial Y^{12}_{A_1}\partial Y^{12}_{B_1}}\bigg|_{\theta=0}\,,
\end{align}
which is indeed exactly the type of coupling considered in (\ref{EffectiveCouplingGauge}). We can see that this coupling just depends on the particular projection $\varphi^{12}$ of the moduli. This would entail that \emph{i.e.} derivatives with respect to, say $\varphi^{34}$ should vanish. Translated into the language of the previous section, this would suggest that that $\mathcal{G}_N$ is independent of $T$. However, as we have seen in Section~\ref{Sect:HolAnom}, this is only correct up to an anomaly. As discussed in \emph{e.g.} \cite{Antoniadis:1993ze} the latter comes about due to the presence of non-analytic terms in the effective action, which (through integrating out auxiliary fields) lead to non-analytic dependencies of the higher derivative couplings. Indeed, as we had seen in Section~\ref{Sect:MockModularForm}, in the holomorphic limit of the amplitude (in which the anomaly is avoided) only short multiplets of the elliptic genus contribute, which is consistent with the $1/2$ BPS nature of the couplings discussed here. The full string amplitude, however, restores holomorphicity of the integrand $\mathcal{G}_N$ (at the cost of modularity) thereby introducing the anomaly as discussed above.
\section{Intermediate Multiplets and the Elliptic Genus}\label{Sect:14BPSamps}
\setcounter{equation}{0}

For $N\geq 1$ the amplitude from  the previous section  receives contributions only from the short multiplets, which contribute to the elliptic genus of $K3$. However, the contribution of intermediate multiplets has dropped from the final expression. 
We want to stress, that the amplitudes considered so far, only involve external states from short (\emph{i.e.} $1/2$ BPS) multiplets. Qualitatively different results should be expected for  amplitudes  involving external states from intermediate or long multiplets. In the following we shall compute these types of amplitudes in two different ways: First we shall consider reducible diagrams, \emph{i.e.} amplitudes with two of the external vertices colliding. Since in the internal channels all kinds of fields are propagating, we will indeed find the elliptic genus of $K3$ in a very particular limit. In order to back up our computation, we will also directly compute these diagrams with massive external legs and show that the two results are in fact identical.
\subsection{Reducible Diagram}
\begin{figure}[ht]
\begin{center}
\input{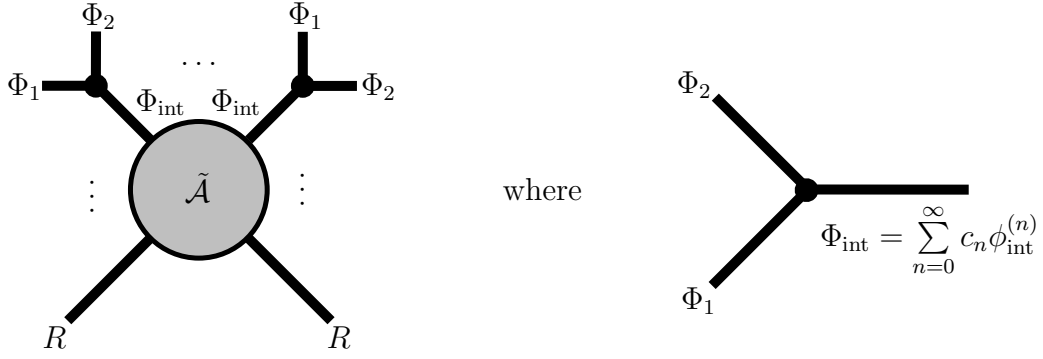}
\caption{\it Reducible diagram with external fields $\Phi_1$ and $\Phi_2$ pairwise colliding. The intermediate channel can be decomposed into a series of fields of different conformal dimensions, which belong to different multiplets. The latter are not necessarily only short ones but might also include $1/4$ BPS multiplets.}
\label{Fig:RedDiagram}
\end{center}
\end{figure}

Our first method of uncovering the elliptic genus of $K3$ will be to consider reducible diagrams. By this we have in mind  diagrams of the type depicted in figure~\ref{Fig:RedDiagram}, where two of the external legs collide and form an intermediate channel. In this channel all kinds of fields, which are allowed by the selection rules, propagate, cf.
the right hand side of figure~\ref{Fig:RedDiagram}. 
By carefully adjusting the way in which the external legs collide, we will see that it is possible to extract the elliptic genus of $K3$. In fact, we will consider two distinct cases, which differ in whether the limit in which the insertion points of two external vertices coincide is singular (and thus needs some proper regularization), or happens to be finite.
\subsubsection{Singular Limit}\label{Sect:SingularLimit}
As first type of reducible amplitude we consider  the coupling
\begin{align}
\Hc_{2N}^{A_1\ldots A_NB_1\ldots B_N}(R_{(+),\mu\nu\rho\tau}R_{(+)}^{\mu\nu\rho\tau})\prod_{i=1}^N\left(\bar{\lambda}_{A_i}\cdot \bar{\lambda}_{B_i}\right)\,,\label{SingLimFieldForm}
\end{align}
with $\bar{\lambda}_A^{\dot{\alpha}}$  the anti--chiral modulini 
associated to the moduli. The relevant vertices can be found in appendix~\ref{App:GenK3Compact}. The coupling function \req{SingLimFieldForm} takes the form:
\begin{align}
&\left\langle \int d^2z_1 V^{(0,0)}_R(h_{11},p_2)\int d^2z_2 V^{(0,0)}_R(h_{\bar{1}\bar{1}},\bar{p}_2)\, \prod_{i=1}^N\int d^2x_i V^{(-1/2,0)}_{\lambda,A_i,{\dot{\alpha}}}(\bar{p}_2^{(i)})\int d^2y_i V^{(+1/2,0),{\dot{\alpha}}}_{\lambda,B_i}(p_2^{(i)})\right\rangle\ .\nonumber
\end{align}
Similar to the computation in the previous section, we can also factorize this amplitude into its bosonic and fermionic part
\begin{align}
\Hc_{2N}=\int d\zeta[N]&\left\langle \prod_{i=1}^Ne^{-\tfrac{\varphi}{2}}S^1\Sigma^{A_i}(x_i)\, e^{\tfrac{\varphi}{2}}S^2\Sigma^{B_i}(y_i)\right\rangle\left\langle\prod_{i=1}^N\bar{\partial} X_3(\bar{x}_i)\partial X_3(y_i)\bar{\partial} X_3(\bar{y_i})\right\rangle\,.\nonumber
\end{align}
We notice that the bosonic part is essentially trivial, since the $X_3$ cannot contract with one another and thus only contribute their zero modes. For the remaining fermionic correlator, we compile all relevant charges in the following table
\begin{center}
\begin{tabular}{|c||c||c||c|c|c|c||c|c|c|c||c|}\hline
&&&&&&&&&&&\\[-10pt]
\textbf{vertex} & \# & \textbf{pos.} & $\phi_1$ & $\phi_2$ & $H_3$ & $H$ & $\tilde{\phi}_1$ & $\tilde{\phi}_2$ & $\tilde{H}_3$ & $\tilde{H}$ & \textbf{picture} \\\hline\hline
&&&&&&&&&&&\\[-10pt]
graviton & $1$ & $z_1$ & $+1$ & $+1$ & $0$ & $0$ & $+1$ & $+1$ & $0$ & $0$ & $(0,0)$\\[2pt]\hline
&&&&&&&&&&&\\[-10pt]
& $1$ & $z_2$ & $-1$ & $-1$ & $0$ & $0$ & $-1$ & $-1$ & $0$ & $0$ & $(0,0)$\\[2pt]\hline\hline
&&&&&&&&&&&\\[-10pt]
modulini & $N$ & $x_i$ & $+\frac{1}{2}$ & $-\frac{1}{2}$ & $+\frac{1}{2}$ & $+\frac{1}{\sqrt{2}}$ & $0$ & $0$ & $0$ & $0$ & $\left(-\frac{1}{2},0\right)$  \\[4pt]\hline
&&&&&&&&&&&\\[-10pt]
& $N$ & $y_i$ & $-\frac{1}{2}$ & $+\frac{1}{2}$ & $-\frac{1}{2}$ & $-\frac{1}{\sqrt{2}}$ & $0$ & $0$ & $0$ & $0$ & $\left(+\frac{1}{2},0\right)$ \\[4pt]\hline
\end{tabular}
\end{center}
We note here that this amplitude will also include a factor reflecting the $SO(n)$-structure of the vertex operators, which will depend on the explicit choice of fields in (\ref{SingLimFieldForm}). Since the modulini $\bar{\lambda}$ are fermionic, this factor must be totally anti-symmetric in all indices. This condition will effectively put an upper limit on $N$, beyond which $\mathcal{H}_{2N}$ will vanish identically. Since this, however, will not turn out important for us in the following, we will drop this gauge-group dependent factor. Using similar methods as in section~\ref{Sect:12BPSandMock} we find for the remaining part of the amplitude
\begin{align}
\Hc_{2N}=\int d\zeta[N]&\left\langle \prod_{i=1}^Ne^{\frac{i}{\sqrt{2}}H}(x_i)\,e^{-\frac{i}{\sqrt{2}}H}(y_i)\right\rangle\ \langle(\partial X_3)^N(\bar{\partial}X_3)^{2N}\rangle\prod_{i<j}E^{1/2}(x_i,x_j)E^{1/2}(y_i,y_j)\nonumber\\
&\times \frac{\vartheta_1\left(z_1-z_2+\tfrac{1}{2}\sum\limits_{i=1}^N(x_i-y_i)\right)\vartheta_1\left(z_1-z_2-\tfrac{1}{2}\sum\limits_{i=1}^N(x_i-y_i)\right)}{E^2(z_1,z_2)\prod_{i,j}^NE^{1/2}(x_i,y_i)}\nonumber
\end{align}
This expression per se is not particularly enlightening since the integration over the world-sheet positions is not very easy to handle. However, we can consider the limits $y_i\to x_i$ for $i=1,\ldots,N$. If we introduce $x_i-y_i=\epsilon_i$ we can compute the following power series expansion 
\begin{align}
\lim_{\epsilon_i\to 0}\Hc_{2N}=&\frac{\left(1+\mathcal{O}(\epsilon_{i})\right)\left(1+\mathcal{O}(\epsilon_{i})\right)}{\prod_{i=1}^N\epsilon_i^{1/2}}\ \langle(\partial X_3)^N(\bar{\partial}X_3)^{2N}\rangle\nonumber\\
&\times\left\langle\prod_{i=1}^N\left(\epsilon_i^{-1/2}+\frac{i\epsilon_i^{1/2}}{\sqrt{2}}\,\partial H(x_i)+\mathcal{O}(\epsilon_i^{3/2})\right)\right\rangle\,.
\end{align}
This expression has obviously a first order pole in each of the $\epsilon_i$. However, minimally subtracting these poles, the (finite) contribution takes the form
\begin{align}
\Hc_{2N}^{\text{finite}}=\left\langle\prod_{i=1}^N\int d^2 x_i J_{K3}(x_i)\right\rangle_{K3}\,\tau_2^{2N+1}(P_L)^N(P_R)^{2N}\,,
\end{align}
which gives rise to the following world--sheet torus integral
\begin{align}
\Hc_{2N}^{\text{finite}}=\int \frac{d^2\tau}{\tau_2}\,\tau_2^{2N}\left[\frac{\partial^N}{\partial z^N}\,\phi_{K3}(\tau,z)\right]_{z=0}\,\sum_{(P_L,P_R)\in\Gamma^{2,2}}P_L^N(P_R)^{2N}\,q^{\frac{1}{2}|P_L|^2}\bar{q}^{\frac{1}{2}|P_R|^2}\,,\label{FullAmplitudeEllipticGenus}
\end{align}
which indeed involves the elliptic genus of $K3$.
\subsubsection{Non-singular Limit}
A different amplitude for which the colliding limit is non--singular is given by taking a slightly different structure of the $SO(n)$-indices. Indeed, the table containing the relevant charges takes the following form
\begin{center}
\begin{tabular}{|c||c||c||c|c|c|c||c|c|c|c||c|}\hline
&&&&&&&&&&&\\[-10pt]
\textbf{vertex} & \# & \textbf{pos.} & $\phi_1$ & $\phi_2$ & $H_3$ & $H$ & $\tilde{\phi}_1$ & $\tilde{\phi}_2$ & $\tilde{H}_3$ & $\tilde{H}$ & \textbf{picture} \\\hline\hline
&&&&&&&&&&&\\[-10pt]
graviton & $1$ & $z_1$ & $+1$ & $+1$ & $0$ & $0$ & $+1$ & $+1$ & $0$ & $0$ & $(0,0)$\\[2pt]\hline
&&&&&&&&&&&\\[-10pt]
& $1$ & $z_2$ & $-1$ & $-1$ & $0$ & $0$ & $-1$ & $-1$ & $0$ & $0$ & $(0,0)$\\[2pt]\hline\hline
&&&&&&&&&&&\\[-10pt]
modulini & $N$ & $x_i$ & $+\frac{1}{2}$ & $-\frac{1}{2}$ & $+\frac{1}{2}$ & $+\frac{1}{\sqrt{2}}$ & $0$ & $0$ & $0$ & $0$ & $\left(+\frac{1}{2},0\right)$  \\[4pt]\hline
&&&&&&&&&&&\\[-10pt]
& $N$ & $y_i$ & $+\frac{1}{2}$ & $-\frac{1}{2}$ & $-\frac{1}{2}$ & $-\frac{1}{\sqrt{2}}$ & $0$ & $0$ & $0$ & $0$ & $\left(-\frac{1}{2},0\right)$ \\[4pt]\hline
&&&&&&&&&&&\\[-10pt]
& $N$ & $u_i$ & $-\frac{1}{2}$ & $+\frac{1}{2}$ & $+\frac{1}{2}$ & $+\frac{1}{\sqrt{2}}$ & $0$ & $0$ & $0$ & $0$ & $\left(+\frac{1}{2},0\right)$ \\[4pt]\hline
&&&&&&&&&&&\\[-10pt]
& $N$ & $v_i$ & $-\frac{1}{2}$ & $+\frac{1}{2}$ & $-\frac{1}{2}$ & $-\frac{1}{\sqrt{2}}$ & $0$ & $0$ & $0$ & $0$ & $\left(-\frac{1}{2},0\right)$ \\[4pt]\hline
\end{tabular}
\end{center}
As before, using this table, we can compute the amplitude in a straight-forward manner
\begin{align}
\Hc_{2N}=&\frac{\vartheta_1\left(a'_1\right)\vartheta_1\left(a'_2\right)\Theta_\Lambda\left(b'\right)}{E^2(z_1,z_2)\prod_{i,j}^NE(x_i,v_i)E(y_i,u_i)}\prod_{i<j}^NE(x_i,x_j) E(y_i,y_j) E(u_i,u_j) E(v_i,v_j)\nonumber\\
&\times  G^{\Lambda}(x_i,y_i)\  \langle(\partial X_3)^N(\bar{\partial}X_3)^{2N}\rangle\,,
\end{align}
where we have used
\begin{align}
a'_1&=z_1-z_2+\sum_{i=1}^N(x_i-y_i+u_i-v_i)\,,&& &&a'_2=z_1-z_2-\sum_{i=1}^N(x_i-y_i+u_i-v_i)\,,\\
b'&=\frac{1}{\sqrt{2}}\sum_{i=1}^N(x_i+y_i-u_i-v_i)\,.
\end{align}
Upon pairwise taking the limits $y_i\to x_i$ and $v_i\to u_i$, the whole amplitude becomes equal to the following correlator of the internal $K3$ theory
\begin{align}
&\Hc_{2N}=\left\langle \prod_{i=1}^Ne^{\sqrt{2}iH}(x_i)\,e^{-\sqrt{2}iH}(u_i)\right\rangle\ \langle(\partial X_3)^{2N}(\bar{\partial}X_3)^{4N}\rangle\,.
\end{align}
However, this is exactly the same correlator already computed in section \ref{Sect:FermCorr}, equation (\ref{K3corrEllGen}), which has indeed been shown to be related to the elliptic genus of $K3$.
\subsection{Massive External States}\label{Sect:EllipticGenus}
We repeat the computation of the previous section by not using the limit of two external vertices colliding, but rather by directly inserting the fields of the intermediate channels. It turns out, that these states correspond to higher excitations from the CFT point of view or to states from intermediate or long multiplets. 
More concretely, we consider closed string states\footnote{We also could consider intermediate states in the first massive level on one side but still massless from the right moving side. According to the level matching condition (\ref{LEVEL}) this requires some non--trivial momenta and windings, \emph{i.e.} the external states would have non-trivial Narain momenta $P_L$ or $P_R$. Since this would lead to unnecessarily complicated amplitudes (except perhaps for particular cases in the moduli space of $\T^2$) we will not further discuss this possibility.}, which are in the first massive level both in the left and right moving sector:
\begin{equation}\label{MASSIVEST}
V^{(-1,-1)}_{M}(p,z,\bar{z})=\ :e^{-\varphi}\,\psi_3\,\partial H(z)\ e^{-\tilde\varphi}\,\tilde{\psi}_3\,\bar{\partial} X_3(\bar{z})\ 
e^{ipX}:\ .
\end{equation}
We refer the reader to appendix \ref{App:MassiveVertex} for further details on massive vertex operators. The state (\ref{MASSIVEST}) corresponds to a massive modulus field. However, we should mention, that from the point of view of the internal CFT living on $K3$, this state is in the right-moving ground state. As we will see, this is the reason why the string amplitudes will finally only be sensitive to the very basic topological information of $K3$, \emph{i.e.} the elliptic genus.
   
To see this in detail, we will consider the following class of amplitudes 
\begin{align}
\left\langle \int d^2z_1 V^{(0,0)}_R(h_{11},p_2)\int d^2z_2 V^{(0,0)}_R(h_{\bar{1}\bar{1}},\bar{p}_2)\prod_{a=1}^N\int d^2x_a\ V^{(-1,-1)}_{M}(p_a,x_a,\bar{x}_a)\prod_{b=1}^NV_{\text{PCO}}(r_b,\bar{r}_b) \right\rangle\label{AmpMassiveVert}
\end{align}
involving $N$ massive fields \req{MASSIVEST}. In order to balance the total ghost--charge we have also inserted a total of $N$ picture changing operators $V_{\text{PCO}}$ at generic world-sheet positions $r_{b=1,\ldots,N}$. We will consider a generic point in the $K3$ moduli space such that the setup under consideration is summarized in the following table
\begin{center}
\begin{tabular}{|c||c||c||c|c|c|c||c|c|c|c||c||c|}\hline
&&&&&&&&&&&&\\[-10pt]
\textbf{vertex} & \# & \textbf{pos.} & $\phi_1$ & $\phi_2$ & $H_3$ & $H$ & $\tilde{\phi}_1$ & $\tilde{\phi}_2$ & $\tilde{H}_3$ & $\tilde{H}$ & \textbf{picture} & \textbf{bosonic}\\\hline\hline
&&&&&&&&&&&&\\[-10pt]
graviton & $1$ & $z_1$ & $+1$ & $+1$ & $0$ & $0$ & $+1$ & $+1$ & $0$ & $0$ & $(0,0)$ &  \\[2pt]\hline
&&&&&&&&&&&&\\[-10pt]
& $1$ & $z_2$ & $-1$ & $-1$ & $0$ & $0$ & $-1$ & $-1$ & $0$ & $0$ & $(0,0)$ &  \\[2pt]\hline\hline
&&&&&&&&&&&&\\[-10pt]
scalar & $N$ & $x_a$ & $0$ & $0$ & $+1$ & $0$ & $0$ & $0$ & $+1$ & $0$ & $\left(-1,-1\right)$  & $\partial H\,\bar{\partial}X_3$ \\[2pt]\hline\hline
&&&&&&&&&&&&\\[-10pt]
PCO & $N$ & $r_b$ & $0$ & $0$ & $-1$ & $0$ & $0$ & $0$ & $-1$ & $0$ & & $\partial X_3\,\bar{\partial}X_3$\\[2pt]\hline
\end{tabular}
\end{center}
Computing the various contractions, the coupling can be written in the form
\begin{align}
\Fc_{N}=&\frac{F_{\Lambda,s}\left(z_1-z_2,z_1-z_2,0\right)}{E^2(z_1,z_2)\tau_2}\cdot  \frac{\tilde{F}_{\Lambda,\tilde{s}}\left(\bar{z}_1-\bar{z}_2,\bar{z}_1-\bar{z}_2,0\right)}{\bar{E}^2(\bar{z}_1,\bar{z}_2)\tau_2}\ \left\langle\prod_{a=1}^N\partial H(x_a)\right\rangle\ \nonumber\\
&\times\left\langle\prod_{a=1}^N\bar{\partial} X_3(\overline x_a)\ \partial X_3(r_a)\ \bar{\partial}X_3(\overline r_a)\right\rangle\,.
\end{align}
Here we have already canceled the contribution of the ghost correlator against the torus contribution on the left- and right moving side. Integrating out the positions of $z_{1,2}$ just results in the first two terms becoming constant. For the remaining expression, however, we notice that the bosonic correlator on $\T^2$ can only contribute through its zero modes and is in particular independent of the world-sheet positions $x_a$. Hence, we are only left with
\begin{align}
\Fc_{N}=\int \frac{d^2\tau}{\tau_2}\,\tau_2^{2N}\left\langle\prod_{a=1}^N\partial H(x_a)\right\rangle_{K3}\  \sum_{(P_L,P_R)\in\Gamma^{2,2}} (P_L)^N(P_R)^{2N} q^{\frac{1}{2}|P_L|^2}\bar{q}^{\frac{1}{2}|P_R|^2}\,.\label{ResultEllipticGenus}
\end{align}
The $K3$ correlator, however, is just the $N$-th derivative of the $K3$ elliptic genus
\begin{align}
\Fc_{N}=\int \frac{d^2\tau}{\tau_2}\,\tau_2^{2N}\left[\frac{\partial^N}{\partial z^N}\,\phi_{K3}(\tau,z)\right]_{z=0}\ \sum_{(P_L,P_R)\in\Gamma^{2,2}} (P_L)^N(P_R)^{2N}\,q^{\frac{1}{2}|P_L|^2}\bar{q}^{\frac{1}{2}|P_R|^2}\,,\label{DerivativeEllGen}
\end{align}
where $\phi_{K3}(\tau,z)$ was introduced in (\ref{EllipticGenusDefinition}). Notice that for all $N$ smaller than the upper limit put in section~\ref{Sect:SingularLimit} we precisely recover the expression (\ref{FullAmplitudeEllipticGenus}), which was the non-singular part of the reducible amplitude depicted in figure~\ref{Fig:RedDiagram}, \emph{i.e.} $\mathcal{F}_N=\mathcal{H}_{2N}$.

As an example and to further check these computations, we have explicitly worked out amplitude (\ref{AmpMassiveVert}) for a torus realization of $K3$ in appendix~\ref{App:Orbifold}.
\section{World--sheet Integral and Igusa Cusp Form $\bm{\chi_{10}}$}\label{Sect:LoopIntegral}
\setcounter{equation}{0}
After the computations of the previous sections we also want to explicitly perform the torus one-loop integrations to obtain some modular form with respect to the T-duality group. To be precise, we will not directly consider (\ref{DerivativeEllGen}), but a somewhat related expression.
First, we define a covariant differential with respect to the modulus $U$ which acts in the following manner on a function $f^{(w)}$ of weight $w$
\begin{align}
\mathcal{D}_{\bar{U}}f^{(w)}=\frac{U-\bar{U}}{2}\left(\frac{\partial}{\partial \bar{U}}-\frac{w}{2(U-\bar{U})}\right)\,f^{(w)}\,,
\end{align}
and where the prefactor takes into account the normalization of the $U$-vertex operators in string theory (see \emph{e.g.} \cite{Lerche:1999ju}) and the second term in the bracket yields the curvature contribution. Using this operator, we can write (\ref{DerivativeEllGen}) in the following manner
{\allowdisplaybreaks
\begin{align}
&\int \frac{d^2\tau}{\tau_2^2}\ \tau_2^{2N+1}\ \left[\frac{\partial^N}{\partial z^N}\,\phi_{K3}(\tau,z)\right]_{z=0}\,\sum_{(P_L,P_R)\in\Gamma^{2,2}}(P_L)^N(P_R)^{2N}\,q^{\frac{1}{2}|P_L|^2}\bar{q}^{\frac{1}{2}|P_R|^2}\nonumber\\
&\hskip1.5cm
=(\mathcal{D}_{\bar{U}})^N\int \frac{d^2\tau}{\tau_2}\,\left[\frac{\partial^N}{\partial z^N}\,\phi_{K3}(\tau,z)\right]_{z=0}\,{\sum_{m_1,m_2\in\mathbb{Z} \atop n_1,n_2\in\mathbb{Z}}}^{\hspace{-0.3cm}'}(\tau_2P_R)^{N}\,q^{\frac{1}{2}|P_L|^2}\bar{q}^{\frac{1}{2}|P_R|^2}\nonumber\\
&\hskip1.5cm=\left(\frac{i}{\pi}\right)^N\left[\prod_{a=0}^{N-1}U_2\left(\frac{\partial}{\partial \bar{U}}+\frac{i(N+2a)}{4U_2}\right)\right]\int \frac{d^2\tau}{\tau_2}\,\left[\frac{\partial^N}{\partial z^N}\,\phi_{K3}(\tau,z)\right]_{z=0}\,\nonumber\\*
&\hskip3cm\times {\sum_{m_1,m_2\in\mathbb{Z} \atop n_1,n_2\in\mathbb{Z}}}^{\hspace{-0.3cm}'}(\tau_2P_R)^{N}\,q^{\frac{1}{2}|P_L|^2}\bar{q}^{\frac{1}{2}|P_R|^2}\ ,
\end{align}}
where the prime at the sum means, that we will not sum over $(m_1,m_2,n_1,n_2)=(0,0,0,0)$. We will from now on focus on the generating functional
\begin{align}
\fun(\lambda ,T,U)&=\int \frac{d^2\tau}{\tau_2}\,\sum_{N=0}^{\infty}\frac{\left(\frac{\lambda }{\sqrt{2T_2U_2}}\right)^N}{N!}\left[\frac{\partial^N}{\partial z^N}\,\phi_{K3}(\tau,z)\right]_{z=0}{\sum_{m_1,m_2\in\mathbb{Z} \atop n_1,n_2\in\mathbb{Z}}}^{\hspace{-0.3cm}'}(\tau_2P_R)^{N}\,q^{\frac{1}{2}|P_L|^2}\bar{q}^{\frac{1}{2}|P_R|^2}=\nonumber\\
&=\int \frac{d^2\tau}{\tau_2}\,{\sum_{m_1,m_2\in\mathbb{Z} \atop n_1,n_2\in\mathbb{Z}}}^{\hspace{-0.3cm}'}\phi_{K3}\left(\tau,\frac{\lambda \tau_2 P_R}{\sqrt{2T_2U_2}}\right)\,q^{\frac{1}{2}|P_L|^2}\bar{q}^{\frac{1}{2}|P_R|^2}\,,
\end{align}
where we have introduced an arbitrary coupling constant $\lambda $ (the factor $\tfrac{1}{\sqrt{2T_2U_2}}$ has been introduced for latter convenience). In the next step we use the fact that $\phi_{K3}$ is a weak Jacobi form an therefore enjoys the following Fourier expansion
\begin{align}
\phi_{K3}(\tau,z)=\sum_{n,\ell}c(n,\ell)q^ne^{2\pi iz\ell}\,,&&\text{with} &&c(n,\ell)=c\left(n-\tfrac{\ell^2}{4}\right)\,,
\end{align}
with the first few coefficients given explicitly
\begin{align}
&c(0)=20\,,&&c(1)=216\,,&&c(2)=1616\,,&&c(3)=8032\,,\\
&c(-1/4)=2\,,&&c(3/4)=-128\,,&&c(7/4)=-1026\,,&&c(11/4)=-5504\,.
\end{align}
With this, we may write for the integral
\begin{align}
\fun(\lambda ,T,U)=\int \frac{d^2\tau}{\tau_2}\,{\sum_{m_1,m_2\in\mathbb{Z} \atop n_1,n_2\in\mathbb{Z}}}^{\hspace{-0.3cm}'}\hspace{0.2cm}\sum_{n,\ell\in\mathbb{Z}}c(n,\ell)\,q^{n-m_2n_1+m_1n_2}\,e^{ \frac{2\pi i\lambda \tau_2\ell P_R}{\sqrt{2T_2U_2}}-2\pi\tau_2 |P_R|^2}\,.\label{PrePoisson}
\end{align}
To explicitly compute the $\tau$-integral (\ref{PrePoisson}) we shall use the methods of orbits, which was first introduced in \cite{Dixon:1990pc,Harvey:1995fq} and further developed in \cite{Borcherds2,Lerche:1999ju,Foerger:1998kw}. 
The whole computation can be found in the appendix~\ref{App:TorusIntegral} and in the following we only present the final result. After putting together the contributions of the three orbits (\ref{ResInteg0}), (\ref{ResIntegND}) and (\ref{ResIntegD}) we obtain the expression
\begin{align}
\fun(\lambda ,T,U)=-4&\ln\left[ Y^{10}\left|e^{2\pi i(T+U+i\lambda )}\prod_{r,n',\ell>0}\left(1-e^{2\pi i(rT+n'U+i\ell \lambda )}\right)^{c(n'r,\ell)}\right|\right]-4\kappa\,,\label{IntegFullRes}
\end{align}
where $Y=T_2U_2-\lambda ^2$. Comparing with \cite{Kawai:1995hy}, we recognize that
\begin{align}
\chi_{10}(T,U;\lambda )=e^{2\pi i(T+U+i\lambda )}\prod_{r,n',\ell>0}\left(1-e^{2\pi i(rT+n'U+i\ell \lambda )}\right)^{c(n'r,\ell)}\,,
\end{align}   
is a product representation of the unique weight 10 Igusa cusp form\footnote{This genus two modular form first appeared in string theory to describe the bosonic two--loop partition function \cite{Moore:1986rh} and in the context of one--loop gauge threshold corrections 
\cite{Mayr:1995rx}.}  of $SO(2,3;\mathbb{Z})\simeq Sp(4,\mathbb{Z})$. We also note that the remaining terms in (\ref{IntegFullRes}) guarantee covariance of the integral under this latter group. Finally, the group $SO(2,3,\mathbb{Z})$ is parametrized by the torus moduli $(T,U)$ (which generically give rise to  $SO(2,2,\mathbb{Z})$) and the additional coupling constant~$\lambda $. 

The inclusion of an additional coupling parameter $\lambda $ as 
an extension of the $T$--duality group  has also been discussed  in 
\cite{Kawai:2000px,Grimm:2007tm}. Furthermore, it has been pointed out in \cite{Hollowood:2003cv}, that the moduli space of mass-deformed six---dimensional $\N=(1,1)$ supersymmetric gauge theories compactified on $\T^2$ enjoys an enhanced symmetry. These theories can be considered  as a mass deformed NS5--brane of type IIA string theory compactified on $\T^2$. Indeed, the two natural $SL(2,\mathbb{Z})$ symmetries of the torus then combine with the mass parameter into an $Sp(4,\mathbb{Z})$ symmetry. It would be  interesting  to see whether there is any connection of these works to the results reported above. 

Moreover, in Ref. \cite{Dijkgraaf:1996it} the modular form $\chi_{10}$ has been identified as partition function counting dyons (dyonic five--brane states) in ${\cal N}=4$  string theory.
In this spirit the parameter $\lambda $ couples to the helicity quantum number $\ell$ of the dyonic state. According to \cite{Dijkgraaf:1996it} by a similar world--sheet integral 
as \req{PrePoisson} the cusp form $\chi_{10}$ is related to the free energy of strings in $D=6$ 
with target--space $K3\times \T^2$ .
The string momenta and windings along $\T^2$  represent the world--volume degrees of freedom on the NS5--brane. The index, which counts these string states is given by the elliptic genus 
\req{EllipticGenusDefinition} of $K3$. It would be also interesting to find a possible connection 
between our one--loop amplitude result 
\req{IntegFullRes} and the computation of the aforementioned  free energy from the partition function of the six--dimensional world--volume string theory on $K3\times\T^2$. 

\section{Conclusions}
In this work we have presented several BPS saturated amplitudes which are linked to the elliptic genus of $K3$. Amplitudes with external fields from  massless short multiplets only capture a particular part of the elliptic genus corresponding to a mock modular form. It has been  argued in the past \cite{Eguchi:2009cq}, that this Appell--Lerch sum is indeed linked to the elliptic genus of massless short multiplets, which thus acts as a confirmation of our computation. Similarly, we have shown that a supersymmetrically related amplitude is in turn linked to the $B_6$ helicity supertrace. It might be interesting to understand whether the mock-modularity of these couplings suggests any further consequences for the space of BPS states. In particular, it might turn out interesting to study this particular coupling under more algebraic aspects, as has been recently done for another class of $\cN=4$ topological amplitudes in \cite{Gaberdiel:2011qu,HP}. 

The full elliptic genus is obtained from amplitudes with external fields sitting in massive multiplets. We present two different methods of computing them: First as the reducible limit of an $1/2$ BPS coupling and secondly directly, using massive external vertex operators. In both cases, we find explicitly derivatives of the elliptic genus of $K3$. It would be interesting to understand whether the appearance of the latter in the explicit string amplitudes signals an action of $\mathbb{M}_{24}$ on the BPS states contributing to these couplings (see \cite{Govindarajan:2010cf,Govindarajan:2009qt} for related discussions). For example, it would be curious to consider the dual amplitudes in heterotic string theory. Since the type II results have a distinct dependence on the $T$-modulus of the $\T^2$ torus (which becomes the heterotic dilaton), one would expect non-perturbative corrections which might suggest that a possible $\mathbb{M}_{24}$ symmetry of the heterotic BPS states could be non-perturbative in nature.

Finally, by introducing a coupling constant $\lambda $, we have introduced a generating functional of the $1/4$ BPS saturated amplitudes. Upon performing explicitly the world-sheet torus integration, we have proven that this generating functional is related to the weight 10 Igusa cusp form of $Sp(4,\mathbb{Z})$. As has first been conjectured in \cite{Dijkgraaf:1996it}, the Fourier coefficients of the latter encode the multiplicities of non-perturbative $1/4$ BPS dyons in toroidally compactified heterotic string theory. It would be interesting to further exploit this connection in the future and obtain a better understanding of the connection between BPS saturated amplitudes and dyon states.
\section*{Acknowledgements}
It is a pleasure to thank Ignatios~Antoniadis, Matthias~Gaberdiel, Thomas~Grimm, Amer Iqbal, Wolfgang Lerche, Kumar Narain, Daniel Persson, Oliver Schlotterer and Tomasz Taylor for many enlightening discussions. SH would like to thank the ICTP Trieste, CERN, Northeastern University and the Simons Center for Geometry and Physics for kind hospitality as well as the organizers of the Simons Summer Workshop in Mathematics and Physics 2011 for creating an inspiring atmosphere during completion of this work. StSt is grateful to CERN for warm hospitality during completion of this work.
\appendix
\section{$\bm{\cN=4}$ Compactifications}
\setcounter{equation}{0}
\subsection[$K3$ Compactifications in Type II String Theory]{$\bm{K3}$ Compactifications in Type II String Theory}\label{App:GenK3Compact}

In this appendix we review some essential points about generic $\cN=(4,4)$ compactifications
following \cite{Antoniadis:2009nv} (see also \cite{Antoniadis:1993ze}). 
As explained in appendix~\ref{App:RepN4}, at a generic point in the moduli space we can bosonize the $U(1)$ current $J_{K3}$ of the internal $\cN=4$ superconformal algebra in terms of a free boson $H$ (see equation (\ref{RepHLattice})). As discussed, the spin structure dependence of the internal contributions will then only enter through shifts and projections of the momentum lattice of $H$, which we will call $\Gamma$. Moreover, the space-time and $\T^2$ torus fermions define an $SO(2)\times SO(2)\times SO(2)$ lattice. We will combine the $SO(2)\times SO(2)$ sublattice stemming from the space-time part with $\Gamma$ to obtain the coset $E_7/SO(8)$ (see \cite{Lerche:1988np,Lechtenfeld:1989be}). We use the same notation as in \cite{Antoniadis:2009nv} and write the characters $\chi_{\Lambda}$ of $E_7$ (in the conjugacy class $\Lambda$) through the branching functions
\begin{align}
\chi_\Lambda(\tau)=\sum_sF_{\Lambda,s}(\tau)\ \chi_s(\tau)\,,
\end{align}
where $\chi_s(\tau)$ are the $SO(8)$ level one characters and $s$ denotes the $4$ conjugacy classes of $SO(8)$ in the spin structure basis. Therefore, the characters of the internal $\cN=4$ SCFT of $K3$ together with the contribution of the space-time free fermions can be written as 
\begin{align}
\sum_{\Lambda} F_{\Lambda,s}(\tau)\ \text{Ch}_{\Lambda}(\tau)\,,\label{GenCharK3}
\end{align}
where $\text{Ch}_{\Lambda}(\tau)$ are the remaining contributions stemming from the $K3$. $\text{Ch}_{\Lambda}(\tau)$ is highly model dependent ({\it i.e.} it depends on the exact position within the $K3$ moduli space) and for most cases is generally not known explicitly. However, as it will turn out crucial for the computations in the main part of this work, $\text{Ch}_{\Lambda}(\tau)$ is independent of the spin-structure. For the one-loop computations, we can furthermore define a more general character $F_{\Lambda,s}(u_1,u_2,v)$ by introducing chemical potentials $u_{1,2}$ coupling to the charges of $SO(2)\times SO(2)$ ({\it i.e.} the space-time part) as well as $v$ coupling to the $H$-charge. As we will see in the main body of the text, $(u_1,u_2,v)$ will be related to the world-sheet positions of the various vertex operators weighted by the corresponding charges via Abel map. Generically, (if needed after an appropriate choice of gauge) the spin structure sum can be computed using 
\begin{align} 
\sum_s F_{\Lambda,s}(u_1,u_2,v)=F_{\Lambda}\left[\tfrac{1}{2}(u_1+u_2+\sqrt{2}v),\tfrac{1}{2}(u_1+u_2-\sqrt{2}v),\tfrac{1}{\sqrt{2}}(u_1-u_2)\right]\,,\label{GenExprSpinStruct}
\end{align}
with the explicit expression
\begin{align}
&F_{\Lambda}(u_1,u_2,v)=\vartheta_1(\tau,u_1)\ \vartheta_1(\tau,u_2)\ \Theta_\Lambda(\tau,v)\,,\label{FLtheta}
\end{align}
where $\Theta_\Lambda$ is the character valued one-loop partition function of level one $SU(2)$
\begin{align}
\Theta_\Lambda(\tau,v)=\sum_{n\in\mathbb{Z}}e^{2\pi i\left(n+\frac{\lambda}{2}\right)^2\tau+2\pi i\sqrt{2}\left(n+\frac{\lambda}{2}\right)v}\,,
\end{align}
where $\lambda=0,1$ corresponding to the $E_7$ conjugacy classes \textbf{1} and \textbf{56} respectively.
\subsection{Vertex Operators}\label{App:Vertex}

To compute scattering amplitudes we need the emission vertex operators for the fields in type II string theory compactified on $K3\times \T^2$. In this appendix we present the relevant expressions. Our conventions follow closely \cite{Antoniadis:1993ze}: we choose a ten-dimensional basis of complex bosonic coordinates $(X_1,\ldots,X_5)$, where $(X_1,X_2)$ denote the four-dimensional space-time, $X_3$ labels the torus $\T^2$  and $(X_4,X_5)$ are the complex coordinates on $K3$. The respective superpartners are called $(\psi_1,\ldots,\psi_5)$ and the right--moving counterparts are denoted with a tilde $(\tilde{\psi}_1,\ldots,\tilde{\psi}_5)$. 
\subsubsection{Internal fields and SCFT}

Both the left-- and right--moving fields give rise to an internal SCFT
specifying an $\Nc=2$ space--time SUSY in each sector. In the following we discuss
one sector.  The space--time supercurrents contain the internal fields $\Si^a,\ov\Si^a$ 
in the Ramond sector of the internal SCFT with $a$  an index of the $SU(2)$ R--symmetry group.  
Their OPEs are given by \cite{Banks:1988yz}
\bea
\Sigma^a(z)\ \ov \Sigma^b(w)&=&(z-w)^{-3/4}\ \delta^{ab}\ I+(z-w)^{1/4}\
\Jc^{ab}(w)+\ldots\ ,\nonumber\\[2mm]
\Sigma^a(z)\ \Sigma^b(w)&=&(z-w)^{-1/4}\ \Psi^{ab}(w)+(z-w)^{3/4}\
\Oc^{ab}(w)+\ldots\ ,
\label{OPE2}
\eaa
with the dimension one currents $\Jc^{ab}$, the dimension $1/2$ operators
$\Psi^{ab}$
and the dimension $3/2$ operators $\Oc^{ab}$.
The internal SCFT splits into two pieces.
One piece is the $c=3$ superconformal algebra, which corresponds
to a torus compactificaton with the two
complex internal fermions $\psi_3=e^{+ iH_3}$ and $\ov\psi_3=e^{-iH_3}$. The second piece
represents a $c=6$ superconformal algebra, which
contains the $SU(2)$ currents $J_{K3}=\fc{i}{\sqrt 2}\pa H,\ J_{K3}^{++}=e^{i\sqrt 2 H}$
and $J_{K3}^{--}=e^{-i\sqrt 2 H}$ \cite{Banks:1988yz}. 
The (internal) Ramond fields $\Si^a$ can be factorized by 
expressing them by the bosonic fields $H_3$ and $H$ as:
\bea
\Si^1&=&e^{\fc{i}{2}H_3}\ \Si_{K3}=e^{\fc{i}{2}H_3}\ e^{\fc{i}{\sqrt2}H}\ ,\nonumber \\
\Si^2&=&e^{\fc{i}{2}H_3}\ \Si_{K3}^\dagger=e^{\fc{i}{2}H_3}\ e^{-\fc{i}{\sqrt2}H}\ .
\label{current}
\eaa
Then, from \req{OPE2} for the internal $K3$ sector we obtain the following OPEs: 
\begin{align}
\Sigma_{K3}(z)\ \Sigma_{K3}^\dagger(w)&=(z-w)^{-1/2}\ I+(z-w)^{1/2}\,J_{K3}(w)+\ldots\,,
\nonumber\\
\Sigma_{K3}(z)\ \Sigma_{K3}(w)&= (z-w)^{1/2}\,J^{++}_{K3}(w)+\ldots\,,\nonumber\\
\Sigma_{K3}^\dagger(z)\ 
\Sigma^\dagger_{K3}(w)&= (z-w)^{1/2}\,J^{--}_{K3}(w)+\ldots\ .\label{OPESUSY}
\end{align}
We refer the reader to appendix \ref{Sect:SCFT} for more details on the underlying superconformal algebras.

\subsubsection{Zero Mass Level}

We start with the  vertex operators from the  massless level, \emph{i.e.} $m_L^2=m_R^2=0$ in \req{MASS} and consider first the NS--NS sector. We  need the vertices of the graviton and 
NS--NS graviphoton fields, which in the $(0,0)$--picture take the form
\begin{align}
V_R^{(0,0)}(h,p,z,\overline z)&=:h_{\mu\nu}\big(\partial X^\mu+i(p\cdot \psi)\psi^\mu\big)
\ \big(\bar{\partial} X^\nu+i(p\cdot \tilde{\psi})\tilde{\psi}^\nu\big)e^{ip\cdot X}:\ ,\label{VertexGraviton}\\
V_F^{(0,0)}(\epsilon,p,z,\overline z)&=:\epsilon_{\mu}\big(\partial X^\mu+i(p\cdot \psi)\psi^\mu\big)\ \big(\bar{\partial} X^3+i(p\cdot \tilde{\psi})\tilde{\psi}^3\big)e^{ip\cdot X}:\ ,\label{VertexNSgauge}
\end{align}
respectively. For the graviton the polarization tensor $h_{\mu\nu}$ is symmetric, traceless and obeys $p^\mu h_{\mu\nu}=0$, while for the graviphoton we have $\epsilon_\mu p^\mu=0$. Furthermore we also  need the vertices of the graviphoton fields from the R--R sector. In the 
$(-1/2,-1/2)$--ghost picture the self--dual and anti--self--dual graviphoton vertices assume the form
\begin{align}
V_{T_{(+)},\mu\nu}^{(-1/2,-1/2),ab}(p,z,\overline z)&=:e^{-\frac{\varphi+\tilde{\varphi}}{2}}\ S^\alpha(\sigma^{\mu\nu})_{\alpha\beta}\ \tilde{S}^\beta\Sigma^a\ \tilde{\Sigma}^b\ e^{ip\cdot X}:\ ,\\
V_{T_{(-)},\mu\nu}^{(-1/2,-1/2),ab}(p,z,\overline z)&=:e^{-\frac{\varphi+\tilde{\varphi}}{2}}\ S_{\dot{\alpha}}(\bar{\sigma}^{\mu\nu})^{\dot{\alpha}\dot{\beta}}\ \tilde{S}_{\dot{\beta}}\ \bar{\Sigma}^a\ \bar{\tilde{\Sigma}}^be^{ip\cdot X}:\ ,
\end{align}
respectively. Finally, we have the vertices of the modulini $\lambda$ stemming from the R--NS sector, which in the $(-1/2,0)$ and $(-1/2,-1)$ picture read
\begin{align}
V_{\lambda}^{(-1/2,0),a\alpha}(p,z,\overline z)&=:e^{-\frac{\varphi}{2}}\ S^\alpha\ \Sigma^a
\ [\bar{\partial} X_3+i(p\tilde \psi)\tilde\psi_3 ]\ e^{ip\cdot X}:\ ,\label{GauginoVertex120}\\
V_{\lambda}^{(-1/2,-1),a\alpha}(p,z,\overline z)&=:e^{-\frac{\varphi}{2}-\tilde{\varphi}}\ 
S^\alpha\ \Sigma^a\ \tilde\psi_3\ e^{ip\cdot X}:\ ,
\end{align}
respectively. Similar expressions with left-- and right--movers exchanged follow for the gauginos from the NS--R sector.

\subsubsection{First Massive Level}\label{App:MassiveVertex}

Let us now move on to massive string oscillator states and consider their vertex operators. 
In the following we shall concentrate on string states with the non--vanishing mass $m_L^2=m_R^2=1$ in \req{MASS}.  Vertices of this type have  been constructed   in $D=10$ 
in \cite{Koh:1987hm}, while some of 
the corresponding $D=4$ states (coupling to the gauge sector) 
can be found in \cite{Feng:2010yx}. 
For $K3\times\T^2$ compactifications in the open string sector in $D=4$ 
there is the massive spin two state $B$
\be
V^{(-1)}_B(z,p)=\ :e^{-\varphi}\ \pa X_3\ \psi_3\ e^{ipX}:
\label{Bmass}
\ee
refering to the (complex) direction along the torus.  In fact, it is straightforward to verify, that this
state is BRST--exact, {\it i.e.}
\be
[\ Q_{BRST},V^{(-1)}_B\ ]=\oint \fc{dz}{2\pi i}\ \ e^\varphi\ \eta\ T_F(z)\ \ 
V^{(-1)}_B(w,p)=0\ , 
\ee
with the fields $\varphi,\eta$ bosonizing the superghost system and the super--current:
\be
T_F=\sum_{j=1}^3 \psi_j\ \pa \bar{X}^j+\bar{\psi}_j\ \pa X^j+T_F^{K3}\ .
\ee
Furthermore in $D=4$ there are the spin zero complex scalar fields \cite{Feng:2010yx}
\be
V^{(-1)}_{\Omega,A}(z,p)=\ :e^{-\varphi}\ \Oc_A\ e^{ipX}:\ ,
\label{Omass}
\ee
with $\Oc_A$  the dimension $3/2$ operators appearing in the OPE \req{OPE2}:
\be
\Oc_A=\left\{
{ e^{iH_3}\ J_{K3}\ =\tfrac{i}{\sqrt 2}\ \psi_3\ \pa H\ ,\atop
  e^{iH_3}\ J^\pm_{K3}\ =\psi_3\ e^{\pm i\sqrt 2 H}\ .}\right.
\ee
Tensoring the two open string states \req{Bmass} and \req{Omass} 
gives rise to a set of massive closed string states with $m_L^2=m_R^2=1$. 
From this set we find the vertex operator 
\req{MASSIVEST}, which may be associated to a (massive) modulus of the internal 
$K3\times\T^2$ compactification.

\section{Harmonic Superspace}\label{App:HarmSuperspace}
\renewcommand{\theequation}{\Alph{section}.\arabic{equation}}
\setcounter{equation}{0}
\subsection{Harmonic Coordinates}
In this appendix we will discuss relevant aspects of four dimensional $\cN=4$ harmonic superspace. Since the R-symmetry group is $SU(4)$, the harmonic coordinates that we introduce will form the coset manifold $SU(4)/H$, where $H$ is a subgroup of $SU(4)$. As has been discussed in \cite{Ivanov:1984ut}, there are several possibilities for the choice of $H$, which are suitable for different physical contexts (see e.g.~\cite{Antoniadis:2007cw} for one particular application). Since we have to deal with several different $1/4$ BPS protected couplings (involving VMs together with the Weyl multiplet) in this work, we will choose $H$ to be the maximal torus to allow us as much flexibility as possible
\begin{align}
\{u^I_i,\bar{u}^I_i\}=\frac{SU(4)}{U(1)\times U(1)\times U(1)}\,.\label{HarmonicCoset}
\end{align}
Our notation here follows~\cite{Andrianopoli:1999vr}, where $i=1,\ldots,4$ is an index of $SU(4)$, while $I=1,2,3,4$ label different sets of charges with respect to $U(1)\times U(1)\times U(1)$. To be explicit, the choice of basis is
\begin{align}
&u_i^1=u_i^{(1,0,1)}\,,&&u_i^2=u_i^{(-1,0,1)}\,,&&u_i^3=u_i^{(0,1,-1)}\,,&&u_i^4=u_i^{(0,-1,-1)}\,.
\end{align}
The $\bar{u}^I_i$ in (\ref{HarmonicCoset}) denote the complex conjugate variables. Since the $u$'s are matrices of $SU(4)$, they satisfy the following relations
\begin{align}
&u_i^Iu_J^i=\delta^I_J\,,&&u_i^Iu_I^j=\delta_i^j\,,&&\epsilon^{ijkl}u^1_iu^2_ju^3_ku^4_l=1\,,
\end{align}
which are sometimes called the \emph{unitarity}- and \emph{unit-determinant} condition, respectively. With these coordinates we can also introduce the projections of the Grassmann variables
\begin{align}
&\theta_I^\alpha=\theta_i^\alpha u_{I}^i\,,&&\text{and} &&\bar{\theta}_{\dot{\alpha}}^{I}=\bar{u}_i^{I}\bar{\theta}_{\dot{\alpha}}^i\,,
\end{align}
such that the integral measure over the full harmonic superspace is given by
\begin{align}
\int d^4\zeta:=\int d^4x\int du\,\prod_{I=1}^4\int d^2\theta_I\int d^2\bar{\theta}^I\,.\label{N4integmeasure}
\end{align}
Here $\int du$ is a harmonic integration, which will just pick the
$SU(4)$ singlet of the integrand. Notice moreover that the integral
involves 16 Grassmann integrals. Superspace couplings which are $1/2$
or $1/4$  BPS protected can be written as integrals with less Grassmann integrations, namely $8$ and $12$ respectively.
\subsection{Linearized On-Shell Superfields}\label{App:LinOnShellMultiplets}
After having introduced the harmonic coordinates, we will now also discuss the superfields living on this space. In four dimensions with $\mathcal{N}=4$ supersymmetry, we have two types of superfields, the supergravity multiplet and VMs. The component expansion of the supergravity multiplet contains the following terms (we concentrate on the bosonic components)
\begin{align}
W=\Phi+\frac{1}{2}(\theta^i\sigma^{\mu\nu}\theta^j)T_{(+),\mu\nu}^{kl}\epsilon_{ijkl}-\frac{1}{12}\epsilon_{ijkl}(\theta^i\sigma^{\mu\nu}\theta^j)(\theta^k\sigma^{\rho\tau}\theta^l)R^{(+)}_{\mu\nu\rho\tau}+\ldots\,.\label{WeylMulti}
\end{align}
Here $\Phi$ is the graviscalar [which in the case of type IIA compactified on $K3\times \T^2$ corresponds to the $T$-modulus of the torus] $T_{(+),\mu\nu}$ is the self-dual part of the graviphoton field-strength tensor and $R^{(+)}_{\mu\nu\rho\tau}$ is the self-dual part of the Riemann tensor. It is important to notice that $W$ is a chiral superfield, {\it i.e.} it satisfies the relations
\begin{align}
&\bar{D}^{\dot{\alpha}}_IW=0\,,&&\forall I=1,2,3,4\,.
\end{align}
For definiteness (and to make connections with \cite{Antoniadis:2007cw}), we can also introduce the following superdescendant of $W$
\begin{align}
K^{\mu\nu}_{IJ}=\epsilon_{IJKL}(\sigma^{\mu\nu})^{\alpha\beta}D_\alpha^{K}D_\beta^{L}W=&T_{(+),ij}^{\mu\nu}u^i_Iu^j_J+(\theta_I\sigma_{\rho\tau}\theta_J)R_{(+)}^{\mu\nu\rho\tau}+\epsilon_{IJKL}(\bar{\theta}^{K}\bar{\sigma}^\tau\sigma^{\mu\nu}\sigma^\rho\bar{\theta}^{L})\partial_\tau\partial_\rho \Phi\nonumber
\end{align}
The second type of superfield is the linearized on-shell VM, which has the following component expansion
\begin{align}
Y^{IJ}=&\varphi^{ij}u_i^{J}u_j^{J}+\epsilon^{IJKL}\theta_{K}^\alpha\lambda_{\alpha i}u^i_{L}+u_i^{I}\bar{\lambda}^{\dot{\alpha},i}\bar{\theta}_{\dot{\alpha}}^J+\epsilon^{IJKL}(\theta_K\sigma^{\mu\nu}\theta_L)F_{(+),\mu\nu}+(\bar{\theta}^I\bar{\sigma}^{\mu\nu}\bar{\theta}^J)F_{(-),\mu\nu}\ldots\,.\label{N4shortVectormulti}
\end{align}
Here, $\varphi^{IJ}=\varphi^{ij}u_i^{J}u_j^{J}$ are six real scalar fields ($\varphi^{ij}=\tfrac{1}{2}\epsilon^{ijkl}\varphi_{kl}$), $\lambda_I=\lambda_iu^i_I$ are the modulini ($\bar{\lambda}^I=\bar{\lambda}^iu_i^I$) and $F_{(\pm),\mu\nu}$ is the (anti-)self-dual part of the gauge field strength. The analyticity properties satisfied by $Y^{IJ}$ can be characterized by
\begin{align}
&D^K_{\alpha}Y^{IJ}=0\,,&&\text{if }K=I\text{ or }K=J\,,\\
&\bar{D}_K^{\dot{\alpha}}Y^{IJ}=0\,,&&\text{if }K\neq I\text{ and }K\neq J\,.
\end{align}
These are just conditions which restrict $Y^{IJ}$ to half the harmonic $\cN=4$ superspace.
\section{World--sheet CFT and Superconformal Algebras}\label{Sect:SCFT}
\setcounter{equation}{0}
In this appendix we present the relevant notation and conventions for the world--sheet CFT of the $\cN=4$ compactifications. As already mentioned, this CFT  decomposes into an $\cN=2$ theory stemming from the torus $\T^2$  and an $\cN=4$ theory from the $K3$. For completeness and in order to set our notation (which mainly follows \cite{Antoniadis:2006mr}) we shall review them briefly in the following, cf. also subsection \ref{SPEC}. 

\subsection[The $\cN=2$ Superconformal Algebra]{The $\bm{\cN=2}$ Superconformal Algebra}\label{App:N2SCA}
In the following we will denote the operators of the $\cN=2$ superconformal algebra (SCA) with a $\T^2$ subscript, in order to emphasize that they come from the torus part of the compactification and in order to distinguish them from the $\cN=4$ part (which will be labelled by a subscript $K3$ as we shall discuss in the following section). The $\cN=2$ SCA has central charge $c=3$ and contains besides the energy momentum tensor $T_{\T^2}$ two  supercurrents $G^\pm_{\T^2}$ which are positively and negatively charged with respect to a $U(1)$ Kac-Moody current $J_{\T^2}$. The conformal weights of these operators are given by $(h_{T_{\T^2}},h_{G^\pm_{\T^2}},h_{J_{\T^2}})=(2,3/2,1)$ and for completeness, we will write the non-trivial operator product expansions (OPE)
\begin{align}
&\hspace{2cm}G_{\T^2}^+(z)G_{\T^2}^-(w)=\frac{6}{(z-w)^3}+\frac{2J_{\T^2}(w)}{(z-w)^2}+\frac{2T_{\T^2}(w)+\partial_wJ_{\T^2}(w)}{z-w}\,,\nonumber\\
&\parbox{16cm}{
\begin{align}
&T_{\T^2}(z)T_{\T^2}(w)=\frac{2T_{\T^2}(w)}{(z-w)^2}+\frac{\partial_wT_{\T^2}(w)}{z-w}\,,&& &&T_{\T^2}(z)G_{\T^2}^{\pm}(w)=\frac{3G_{\T^2}^\pm(w)}{2(z-w)^2}+\frac{\partial_wG_{\T^2}^\pm(w)}{z-w}\,,\nonumber\\
&T_{\T^2}(z)J_{\T^2}(w)=\frac{J_{\T^2}(w)}{(z-w)^2}+\frac{\partial_wJ_{\T^2}(w)}{z-w}\,, && &&J_{\T^2}(z)G_{\T^2}^\pm(w)=\pm\frac{G_{\T^2}^\pm(w)}{z-w}\,,\nonumber\\
&J_{\T^2}(z)J_{\T^2}(w)=\frac{2}{(z-w)^2}\,.\nonumber
\end{align}}\nonumber
\end{align}

\subsection[The $\cN=4$ Superconformal Algebra]{The $\bm{\cN=4}$ Superconformal Algebra}\label{App:N4SCA}
In distinction to the $\cN=2$ theory, the $\cN=4$ SCA has central charge $c=6$ and contains besides the energy momentum tensor $T_{K3}$ two doublets of supercurrents $(G^+_{K3},\tilde{G}^-_{K3})$ and $(\tilde{G}^+_{K3},G^-_{K3})$, which transform under an $SU(2)$ Kac-Moody current algebra formed by $(J^{\pm\pm}_{K3},J_{K3})$. The conformal weights are respectively given by $(h_{T_{K3}},h_{G^\pm_{K3}},h_{\tilde{G}^\pm_{K3}},h_{J^{\pm\pm}_{K3}},h_{J_{K3}})=(2,3/2,3/2,1,1)$. The non-trivial OPEs of these objects are given by
{\allowdisplaybreaks
\begin{align}
&G^+_{K3} (z)G^-_{K3} (0)\sim\frac{J_{K3} (0)}{z^2}-\frac{T^B_{K3} (0)-\frac{1}{2}\partial J_{K3} (0)}{z}\,,&& &&J^{--}_{K3}(z)G^+_{K3} (0)\sim\frac{\tilde{G}^-_{K3} (0)}{z}
\nonumber\\
&\tilde{G}^+_{K3} (z)\tilde{G}^-_{K3} (0)\sim\frac{J_{K3} (0)}{z^2}-\frac{T^B_{K3} (0)-\frac{1}{2}
\partial J_{K3} (0)}{z}\,,&& &&J ^{++}_{K3}(z)\tilde{G} ^-_{K3}(0)
\sim-\frac{G^+_{K3} (0)}{z}\,,\nonumber\\
&\tilde{G}^+_{K3} (z)G^+_{K3} (0)\sim\frac{2J^{++}_{K3} (0)}{z^2}+\frac{\partial J^{++}_{K3} (0)}{z}\,,&& &&J_{K3}^{++}(z)G^-_{K3} (0)\sim \frac{\tilde{G}^+_{K3} (0)}{z}
\nonumber\\
&\tilde{G}^-_{K3} (z)G^-_{K3} (0)\sim\frac{2J^{--}_{K3} (0)}{z^2}+\frac{\partial J^{--}_{K3} (0)}{z}\,,&& &&J^{--}_{K3}(z)
\tilde{G}^+_{K3} (0)\sim-\frac{G^-_{K3} (0)}{z}
\nonumber
\end{align}
}
while for any operator $O^q $ of $U(1)$ charge $q$, one has:
\begin{align}
J_{K3} (z)O^q (0)\sim q\frac{O^q (0)}{z}\,.\label{U1algebra}
\end{align}
Oftentimes it will be useful to combine the $\cN=4$ supercurrents into $SU(2)$-doublets in the following manner
\begin{align}
&G^+_{K3,i}\equiv\left(\begin{array}{c}\tilde{G}^+_{K3} \\ G^+_{K3}\end{array}\right),&&\text{and} &&G^-_{K3,i}\equiv \left(\begin{array}{c}G^-_{K3} \\ -\tilde{G}^-_{K3}\end{array}\right)\,.\label{CovarSupercurrents}
\end{align}
which we can furthermore project with the harmonic variables defined in appendix~\ref{App:HarmSuperspace}, \emph{i.e.} $G^+_{K3,\pm}:=G^+_{K3,i}\bar{u}^i_\pm$ and $G^-_{K3,\pm}:=G^-_{K3,i}\bar{u}^i_\pm$.
\subsection{Representations}\label{App:RepN4}
For the computations in the main body of the text, it will be important for us to have an explicit representation of the world-sheet superconformal algebra, which realizes the above OPE relations. An explicit representation of the $\cN=2$ SCA can be written in terms of a free complex boson $X_3$ and fermion $\psi_3=e^{+ iH_3}$ (and $\ov\psi_3=e^{-iH_3}$) living on $\T^2$, \emph{i.e.}
\begin{align}
&T_{\T^2}=\frac{1}{2}\psi_3{\leftrightarrow\atop\displaystyle{\partial\atop~}}\bar{\psi}_3+\partial X_3\partial \bar{X}_3,&&G^-_{\T^2}=\bar{\psi}_3\partial X_3, &&G^{+}_{\T^2}=\psi_3\partial\bar{X}_3, &&J_{\T^2}=\psi_3\bar{\psi}_3\ .
\end{align}
Concerning the $\cN=4$ algebra, for most of our computations it will turn out to be sufficient to have a representation of the $SU(2)$ current algebra $(J_{K3},J^{\pm\pm}_{K3})$. At a generic point in the $K3$ moduli space, the latter can be bosonized in terms of a free boson $H$ such that
\begin{align}
&J_{K3}=\frac{i}{\sqrt{2}} \partial H\,,&&J^{\pm\pm}_{K3}=e^{\pm i\sqrt{2}H}\,,&&G^{\pm}_{K3,i}=e^{\pm \frac{i}{\sqrt{2}}H}\hat{G}_{K3,i}\,,\label{RepHLattice}
\end{align}
where $\hat{G}_{K3,i}$ has conformal dimension $5/4$ and non-singular OPE with the scalar field $H$. For us it will be important that in the computation of amplitudes the spin structure dependence enters only through the projections and shifts in the $U(1)$ charge lattice of $J_{K3}$, which in turn is given by the momentum lattice of $H$. Therefore, in the internal $\cN=4$ theory, only the partition function and correlation functions of $H$ depend on the spin structure.

For several applications we will nevertheless find it useful to have a representation of the full $\cN=4$ SCA. Here we will give explicit expressions for torus orbifold models of $K3$, in terms of free bosons $X_{4,5}$ and fermions $\psi_{4,5}$
\begin{align}
T_{K3} =\partial X_4\partial{\bar X}_4+\partial X_5\partial{\bar X}_5+
{1\over 2}(\psi_4{\leftrightarrow\atop\displaystyle{\partial\atop~}}{\bar\psi}_4+
\psi_5{\leftrightarrow\atop\displaystyle{\partial\atop~}}{\bar\psi}_5)\ ,
\end{align}
${}$\vspace{-0.8cm}
\begin{align}
&J_{K3} =\psi_4\bar{\psi}_4+\psi_5\bar{\psi}_5, && J^{++}_{K3} =\psi_4\psi_5, &&J^{--}_{K3} 
=\bar{\psi}_4\bar{\psi}_5\,,
\label{N4curs}
\end{align}
\begin{align}
&G^+_{K3} =\psi_4\partial \bar{X}_4+\psi_5\partial\bar{X}_5, &&\tilde{G}^+_{K3} =-\psi_5\partial X_4+\psi_4\partial X_5\,,\label{K3scurrent1}\\
&G^{-}_{K3} =\bar{\psi}_4\partial X_4+\bar{\psi}_5\partial X_5, &&\tilde{G}^-_{K3} =-\bar{\psi}_5\partial \bar{X}_4+\bar{\psi}_4\partial \bar{X}_5\,.
\label{K3scurrent2}
\end{align}

\section{Example: Orbifold Compactification}\label{App:Orbifold}
\setcounter{equation}{0}
For concreteness we will supplement the generic computations of section~\ref{Sect:EllipticGenus} by a particular example, namely an orbifold realization of $K3$. At this particular point in moduli space, the $U(1)$ current $J_{K3}=i\sqrt{2}\partial H$ in the massive scalar vertices (\ref{MASSIVEST}) can be  represented as the fermion bilinear term (\ref{N4curs}). Indeed, it is sufficient to consider the term $\psi_4\bar{\psi}_4$, as the similar expression  $\psi_5\bar{\psi}_5$ gives the same result. Therefore, the setup we consider can be summarized in the following table
\begin{center}
\begin{tabular}{|c||c||c||c|c|c|c|c||c|c|c|c|c||c||c|}\hline
&&&&&&&&&&&&&&\\[-10pt]
\textbf{vert.} & \# & \textbf{p.} & $\phi_1$ & $\phi_2$ & $H_3$ & $H_4$ & $H_5$ & $\tilde{\phi}_1$ & $\tilde{\phi}_2$ & $\tilde{H}_3$ & $\tilde{H}_4$ & $\tilde{H}_5$ & \textbf{pict.} & \textbf{bos.}\\\hline\hline
&&&&&&&&&&&&&&\\[-10pt]
grav. & $1$ & $z_1$ & $+1$ & $+1$ & $0$  & $0$ & $0$ & $+1$ & $+1$ & $0$  & $0$ & $0$ & $(0,0)$ &  \\[2pt]\hline
&&&&&&&&&&&&&&\\[-10pt]
& $1$ & $z_2$ & $-1$ & $-1$ & $0$ & $0$  & $0$ & $-1$ & $-1$ & $0$ & $0$ & $0$ & $(0,0)$ &  \\[2pt]\hline\hline
&&&&&&&&&&&&&&\\[-10pt]
scalar & $N$ & $x_a$ & $0$ & $0$ & $+1$ & $\pm 1$  & $0$ & $0$ & $0$ & $+1$ & $0$  & $0$ & \!$\left(-1,-1\right)$\!  & $\bar{\partial}X_3$ \\[2pt]\hline\hline
&&&&&&&&&&&&&&\\[-10pt]
PCO & $N$ & $r_a$ & $0$ & $0$ & $-1$ & $0$  & $0$ & $0$ & $0$ & $-1$ & $0$  & $0$ & & $\partial X_3\bar{\partial}X_3$\\[2pt]\hline
\end{tabular}
\end{center}
Here, for computational convenience, we will use the following expression for the scalar vertex at position $x_a$ 
\begin{align}
V^{(-1,-1)}_{\text{scal}}(x_a)=\lim_{\epsilon_a\to 0}:e^{i(H_3+H_4)}(x_a)e^{-iH_4}(x_a-\epsilon_a):\,.\label{VertexEpsilon}
\end{align}
Notice in particular the normal ordering which makes sure that the limit $\epsilon_a\to0$ is non-singular. Computing the contractions in the usual manner we obtain
\begin{align}
\Fc_{N}=&\lim_{\epsilon_a\to 0}\frac{\vartheta_s\!(z_1-z_2)^2\vartheta_s\!\left(\sum_{a}(x_a-r_a)\right)\vartheta_{h,s}\!(\sum_{a}\epsilon_a)\vartheta_{-h,s}\!(0)}{\vartheta_s\!\left(\sum_{a}(x_a-r_a)+2\Delta\right)E^2(z_1,z_2)\prod_{a}E(x_a,x_a-\epsilon_a)}\nonumber\\
\times&\frac{\bar{\vartheta}_s\!(\bar{z}_1-\bar{z}_2)^2\bar{\vartheta}_s\!\left(\sum_{a}(\bar{x}_a-\bar{r}_a)\right)\bar{\vartheta}_{h,s}\!(0)\vartheta_{-h,s}\!(0)}{\bar{\vartheta}_s\!\left(\sum_{a}(\bar{x}_a-\bar{r}_a)+2\Delta\right)\bar{E}^2(\bar{z}_1,\bar{z}_2)}\ \left\langle\prod_{a=1}^N\bar{\partial} X_3(\bar{x}_a) \partial X_3(r_a)\bar{\partial}X_3(\bar{r}_a)\right\rangle\,.
\end{align}
In order to be able to perform the sum over spin structures, we choose the following gauge fixing condition
\begin{align}
\sum_{a=1}^Nr_a=\sum_{a=1}^Nx_a-z_1+z_2+2\Delta\,,
\end{align}
such that we obtain the result
\begin{align}
\Fc_{N}=&\lim_{\epsilon_a\to 0}\frac{\vartheta\!\left(z_1-z_2+\sum_a\tfrac{\epsilon_a}{2}-\Delta\right)\vartheta\!\left(z_2-z_1+\sum_a\tfrac{\epsilon_a}{2}-\Delta\right)\vartheta_{h}\!\left(\sum_{a}\tfrac{\epsilon_a}{2}-\Delta\right)\vartheta_{-h}\!\left(\sum_{a}\tfrac{\epsilon_a}{2}-\Delta\right)}{E^2(z_1,z_2)\prod_{a}E(x_a,x_a-\epsilon_a)}\nonumber\\
\times&\frac{\bar{\vartheta}(\bar{z}_1-\bar{z}_2-\Delta)\bar{\vartheta}(\bar{z}_2-\bar{z}_1-\Delta)\bar{\vartheta}_{h}\!(0)\vartheta_{-h}\!(0)}{\bar{E}^2(\bar{z}_1,\bar{z}_2)}\ \left\langle\prod_{a=1}^N\bar{\partial} X_3(\bar{x}_a) \partial X_3(r_a)\bar{\partial}X_3(\bar{r}_a)\right\rangle\,.
\end{align}
Noticing that the bosonic correlator can only contribute zero modes and is thus independent of the positions $x_a$, we can write
\begin{align}
\Fc_{N}=&\lim_{\epsilon\to 0}\frac{\vartheta\!\left(z_1-z_2+\sum_a\tfrac{\epsilon_a}{2}-\Delta\right)\vartheta\!\left(z_2-z_1+\sum_a\tfrac{\epsilon_a}{2}-\Delta\right)}{E^2(z_1,z_2)\prod_{a}E^{1/2}(x_a,x_a-\epsilon_a)}\ \left\langle( \partial X_3)^n(\bar{\partial} X_3)^{2n}\right\rangle\nonumber\\
\times&\left\langle\prod_{a=1}^Ne^{\frac{i}{2}(H_4+H_5)}(x_a)e^{-\frac{i}{2}(H_4+H_5)}(x_a+\epsilon_a)\right\rangle\,.
\end{align}
Expanding this expression in power series in $\epsilon_a$, we find
\begin{align}
\Fc_{N}=&\lim_{\epsilon\to 0}\frac{\left(1+o(\epsilon_a)\right)\left(1+o(\epsilon_a)\right)}{\prod_a\left(\epsilon_a^{1/2}+\mathcal{O}\left(\epsilon_a^{3/2}\right)\right)}\ \left\langle( \partial X_3)^n(\bar{\partial} X_3)^{2n}\right\rangle\nonumber\\
\times&\left\langle\prod_{a=1}^N\left(\tfrac{1}{\epsilon_a^{1/2}}+\epsilon_a^{1/2}(\psi_4\bar{\psi_4}+\psi_5\bar{\psi_5})+\mathcal{O}\left(\epsilon_a^{3/2}\right)\right)\right\rangle\,.
\end{align}
The leading order pole is in fact canceled due to the normal ordering of the vertex in (\ref{VertexEpsilon}). Performing also the integration over $z_{1,2}$ we obtain 
\begin{align}
\Fc_{N}=&\left\langle\prod_{a=1}^NJ_{K3}(x_a)\right\rangle\left\langle( \partial X_3)^N(\bar{\partial} X_3)^{2N}\right\rangle\,,
\end{align}
which is equivalent to the more general result obtained in (\ref{ResultEllipticGenus}).
\section{World--sheet Torus Integral}\label{App:TorusIntegral}
\setcounter{equation}{0}
In this appendix we shall explicitly evaluate the integral (\ref{PrePoisson}). The first step is to perform a Poisson resummation on the integers $m_1,m_2$. This is possible in a straight-forward way since these variables appear at most quadratic in the exponent of (\ref{PrePoisson}). The result is
\begin{align}
\fun(\lambda ,T,U)=T_2\int \frac{d^2\tau}{\tau_2^2}\,\sum_{k_{1,2},n_{1,2}\in\mathbb{Z}\atop{n,\ell\in\mathbb{Z}}}\,c(n,\ell)q^{n}\,e^{-\frac{\pi T_2}{U_2\tau_2}|\mathcal{A}|^2-2\pi iT\text{det}A+\frac{\pi i\ell \lambda }{U_2}\mathcal{A}}\,,\label{START}
\end{align}
where we have introduced 
\begin{align}
&A=\left(\begin{array}{cc}n_1 & -k_2 \\ n_2 & k_1\end{array}\right)\,,&&\text{and} &&\mathcal{A}=(1,U)\, A\,\big(^{\tau}_{\hspace{0.01cm}1}\big)\,.
\end{align}
The evaluation of (\ref{START}) is devoted to the appendix \ref{App:TorusIntegral}.
In (\ref{START}) the sum over the integers $k_1,k_2,n_1,n_2$ can be interpreted as a summation over all $2\times 2$ matrices with integer entries. As has been discussed in \cite{Dixon:1990pc}, using modular covariance of the integrand, we can rewrite this sum in terms of inequivalent orbits of $\Gamma=SL(2,\mathbb{Z})$
\begin{align}
\fun(\lambda ,T,U)=\fun^{(0)}(\lambda ,T,U)+\fun^{(\text{ND})}(\lambda ,T,U)+\fun^{(\text{D})}(\lambda ,T,U)\,.\label{TorusContributions}
\end{align}
Indeed, the three cases  to be considered in more detail are distinguished by their choice of representative matrix $A_0$. To  evaluate the three distinct contributions in (\ref{TorusContributions}) we mainly follow a similar computation performed in \cite{David:2006ji} (see also \cite{Gaberdiel:2011qu}).
\begin{itemize}
\item zero orbit: the zero orbit consists of the $SL(2,\mathbb{Z})$ orbit of the representative matrix $A_0=0$. In this case, the integral over the fundamental domain $\mathbb{H}/\Gamma$ of $\Gamma=SL(2,\mathbb{Z})$ can be performed by elementary methods and yields the result
\cite{Lerche:1987qk}

\begin{align}
\fun^{(0)}(\lambda ,T,U)=T_2\int_{\mathbb{H}/\Gamma} \frac{d^2\tau}{\tau_2^2}\,\phi_{K3}(\tau,0)=8\pi T_2\,.\label{ResInteg0}
\end{align}
\item non-degenerate orbit: we choose the representative matrix
\begin{align}
&A=\left(\begin{array}{cc}r & j \\ 0 & p \end{array}\right)\,,&&\text{with} &&\begin{array}{l}p\in\mathbb{Z}\neq 0 \\ r>j\geq 0\end{array}\,.
\end{align}
As has been explained in \cite{Dixon:1990pc}, using covariance of the integrand under $SL(2,\mathbb{Z})$, the integration region can be extended to the (double cover of the) upper half-plane $\mathbb{H}$. To perform the integration over $\tau_1\in(-\infty,\infty)$, we choose the following coordinate transformation
\begin{align}
&\tau'_1=r\tau_1+j+pU_1\,,&&\text{such that} &&\mathcal{A}=\tau'_1+i(pU_2+r\tau_2)\,.
\end{align}
We notice, that after this change of variables, the only dependence on $j$ comes from the factor $q^n=e^{-2\pi n\tau_2+\frac{2\pi in}{r}(\tau'_1-j-pU_1)}$. The summation over $0\leq j\leq r-1$, yields either a factor of $r$ when $n$ is a multiple of $r$ and vanishes else. Upon introducing $n=n'r$ where $n'\geq 0$ we can write
\begin{align}
\fun^{(\text{ND})}(\lambda ,T,U)=2T_2\int_{\mathbb{H}} \frac{d\tau'_1d\tau_2}{\tau_2^2}\,&\sum_{{r=1}\atop {n'=0}}^\infty\sum_{\ell,p\in\mathbb{Z}}\,c(n'r,\ell)e^{-2\pi n'r\tau_2+2\pi in'(\tau'_1-pU_1)}\nonumber\\
\times& e^{-\frac{\pi T_2}{U_2\tau_2}[{\tau'_1}^2+(pU_2+r\tau_2)^2]-2\pi iTrp+\frac{\pi i\ell \lambda }{U_2}[\tau'_1+i(pU_2+r\tau_2)]}\,.
\end{align}
The next step is to perform the integration over $\tau'_1$, which is straight-forward, since it is just a Gaussian integral
\begin{align}
&\fun^{(\text{ND})}(\lambda ,T,U)=\nonumber\\
&=2\sqrt{T_2U_2}\int_{0}^\infty \frac{d\tau_2}{\tau_2^{3/2}}\,&\sum_{{r=1}\atop {n'=0}}^\infty\sum_{\ell,p\in\mathbb{Z}}\,c(n'r,\ell)e^{-\frac{\pi p^2T_2U_2}{\tau_2}-\frac{\pi \tau_2(\lambda \ell+2rT_2+2U_2n')^2}{4T_2U_2}-p\pi (\lambda \ell+2irT_1+2in'U_1)}\,.\nonumber
\end{align}
The integral over $\tau_2$ is of Bessel type and can also be evaluated explicitly
\begin{align}
\fun^{(\text{ND})}(\lambda ,T,U)=-&\sum_{p=1}^\infty\sum_{r,n',\ell}\,\frac{2}{p}\ c(n'r,\ell)\ 
\left[e^{2\pi i(rT+n'U+i\ell \lambda )}+e^{-2\pi i(r\bar{T}+n'\bar{U}-i\ell \lambda )}\right]\,.
\end{align}
Finally, the sum over $p$ can be performed using $\sum_{p=1}^\infty\tfrac{x^p}{p}=\ln(1-x)$, to give
\begin{align}
\fun^{(\text{ND})}(\lambda ,T,U)=-&\,\ln\prod_{r,n',\ell}\left|1-e^{2\pi i(rT+n'U+i\ell \lambda )}\right|^{4c(n'r,\ell)}\,.\label{ResIntegND}
\end{align}
\item degenerate orbit: we choose the representative matrix to be of the form 
\begin{align}
&A_0=\left(\begin{array}{cc} 0 & j \\ 0 & p \end{array}\right)\,,&&\text{with} &&\begin{array}{l}j,p\in\mathbb{Z} \\ (j,p)\neq (0,0) \end{array}
\end{align}
The integration region for this contribution is extended to the semi-infinite strip $\mathbb{S}:\,(\tau_1,\tau_2)\in[-1/2,1/2)\times [0,\infty)$. The only $\tau_1$ dependence of the integrand is encoded in the term $q^n$, which thus forces $n=0$. The integral over $\tau_2$ is standard and can be performed in a straight-forward manner. Moreover, since $c(n=0,\ell)=0$ for all $\ell\notin\{\pm1,0\}$, we find the following contribution
\begin{align}
\fun^{(\text{ND})}(\lambda ,T,U)=\frac{U_2}{\pi}\sum_{{p,j}}\sum_{l=-1}^1\,c(0,\ell)\,\frac{e^{\frac{\pi i\lambda \ell}{U_2}(j+pU)}}{|j+pU|^2}\,.
\end{align}
Expressions of this type have already been considered before in \cite{Foerger:1998kw,David:2006ji} and we can therefore simply state the results. The contribution of $\ell=0$ is given by
\begin{align}
\fun^{(\text{ND})}_{\ell=0}(\lambda ,T,U)=20\left(\frac{\pi U_2}{3}-\ln(T_2U_2-\lambda ^2)-\kappa\right)-\ln\prod_{n=1}^{\infty}\left|1-e^{2\pi i n'U }\right|^{4c(0,0)}\,,\nonumber
\end{align}
where $\kappa=\ln\left(\frac{8\pi}{3\sqrt3}e^{1-\gamma_E}\right)$ and $\gamma_E$ is the Euler-Mascheroni constant. Similarly, we can immediately state the result for the $\ell=\pm1$ contribution
\begin{align}
\fun^{(\text{ND})}_{\ell=\pm1}(\lambda ,T,U)=&4\pi\left(\frac{\lambda ^2}{U_2}+\lambda +\frac{U_2}{6}\right)-8\pi \lambda -\ln\prod_{{n'>0}\atop \ell=-1,1}\left|1-e^{2\pi i (n'U+i\ell \lambda )}\right|^{4c(0,\ell)}-\nonumber\\
&-\ln\left|1-e^{-2\pi \lambda }\right|^{4c(0,1)}\,.\label{ResIntegD}
\end{align}
\end{itemize}
\break

\end{document}